\documentclass[english,american,10pt,letterpaper]{article}
\usepackage[T1]{fontenc}
\usepackage{geometry}
\usepackage{color}
\usepackage{babel}
\usepackage{array}
\usepackage{prettyref}
\usepackage{float}
\usepackage{multirow}
\usepackage{amsmath}
\usepackage{amsthm}
\usepackage{amssymb}
\usepackage{graphicx}
\usepackage{setspace}
\usepackage[numbers]{natbib}
\usepackage[unicode=true,
 bookmarks=true,bookmarksnumbered=false,bookmarksopen=true,bookmarksopenlevel=2,
 breaklinks=false,pdfborder={0 0 0},pdfborderstyle={},backref=false,colorlinks=false]
 {hyperref}
\hypersetup{pdftitle={Correlated activity of periodically driven binary networks},
 pdfauthor={Tobias Kühn, Moritz Helias},
 unicode=true, bookmarks=true,bookmarksnumbered=false,bookmarksopen=true, breaklinks=false,pdfborder={0 0 0},backref=false,colorlinks=false}

\makeatletter

\providecommand{\tabularnewline}{\\}

\usepackage{ifthen}
\newboolean{isarxiv}

\usepackage{changepage}

\usepackage[utf8]{inputenc}

\usepackage{textcomp,marvosym}

\usepackage{fixltx2e}

\usepackage{amsmath,amssymb}

\usepackage{cite}

\usepackage{nameref}

\usepackage[right]{lineno}

\usepackage{microtype}
\DisableLigatures[f]{encoding = *, family = * }

\usepackage{rotating}

\usepackage{setspace} 
\raggedright
\usepackage[aboveskip=3pt,labelfont=bf,labelsep=period,justification=raggedright,singlelinecheck=off]{caption}

\makeatletter
\renewcommand{\@biblabel}[1]{\quad#1.}
\makeatother

\date{}

\usepackage{epstopdf}
\ifthenelse{\boolean{isarxiv}}{\newsavebox\ploslogo \savebox{\ploslogo}{\includegraphics[width=2in]{PLOS-submission.eps}}}{}

\usepackage{float}

\usepackage{lastpage,fancyhdr,graphicx}
\usepackage{epstopdf}
\fancyheadoffset[L]{2.25in}
\fancyfootoffset[L]{2.25in}
\usepackage{tabularx} 
\usepackage{multirow}
\usepackage{colortbl}
\usepackage{dsfont}

\usepackage{prettyref}

\newrefformat{cap}{\hyperref[#1]{Fig~\ref{#1}}}
\newrefformat{fig}{\hyperref[#1]{Fig~\ref{#1}}}
\newrefformat{tab}{\hyperref[#1]{Table ~\ref{#1}}}
\newrefformat{sec}{\hyperref[#1]{Sec.~\ref{#1}}}
\newrefformat{sub}{\hyperref[#1]{Sec.~\ref{#1}}}
\newrefformat{eq}{\hyperref[#1]{Eq~(\ref{#1})}}

\newrefformat{appfig}{\hyperref[#1]{App-Fig~\ref{#1}}}
\newrefformat{appeq}{\hyperref[#1]{App-Eq~(\ref{#1})}}
\newrefformat{appsec}{\hyperref[#1]{App-Sec.~\ref{#1}}}
\newrefformat{appsubsec}{\hyperref[#1]{App-Sec~\ref{#1}}}

\usepackage{braket}
\usepackage{bbm}
\definecolor{orange}{rgb}{1.0,0.8,0.}

\usepackage{xr}
\usepackage[OT1]{fontenc}

\usepackage{etoolbox}
\makeatletter
\patchcmd{\@startsection}{\@ssect{#3}{#4}{#5}{#6}}{\@dblarg{\@sect{#1}{\@m}{#3}{#4}{#5}{#6}}}{}{\PackageError{fix-unnumbered-sections}{Unable to patch \string\@startsection; are you using a non-standard document class?}\@ehd}
\makeatother

\makeatother

\begin{document}
\selectlanguage{english}%
\global\long\def\c{\bar{c}}
\global\long\def\a{\bar{a}}
\global\long\def\m{\bar{m}}
\global\long\def\hext{h_{\text{ext}}}
\global\long\def\St{S\text{-terms}}
\global\long\def\Sht{S_{h}\text{-term}}
\global\long\def\Smt{S_{m}\text{-term}}
\global\long\def\Sinhom{\text{susceptibility terms}}
\global\long\def\Shinhom{\text{direct drive}}
\global\long\def\Sminhom{\text{recurrent drive}}
\global\long\def\diag{\mathrm{diag}}
\foreignlanguage{american}{}\global\long\def\ms{\,\mathrm{ms}}
\foreignlanguage{american}{}\global\long\def\Hz{\,\mathrm{Hz}}
\global\long\def\ainhom{\text{modulated-autocorrelations-drive}}
\global\long\def\at{a\text{-term}}
\global\long\def\mE{m_{\mathrm{exc}}}
\global\long\def\mI{m_{\mathrm{inh}}}
\global\long\def\Gott{\text{jenes höhere Wesen, das wir verehren}}

\selectlanguage{american}%
\let\oldnameref\nameref \renewcommand{\nameref}[1]{\textit{``\oldnameref{#1}''}}

\selectlanguage{english}%
\definecolor{orange}{rgb}{1.0,0.8,0.}

\selectlanguage{american}%
\setboolean{isarxiv}{true}

\vspace*{0.35in}

\begin{flushleft}

{
\Large \textbf
\newline
{Locking of correlated neural activity to ongoing oscillations} 
\addcontentsline{toc}{section}{Title}
}\newline
\\
Tobias K\"uhn \textsuperscript{1,*}, Moritz Helias \textsuperscript{1,2}
\\
\bigskip
{\bf 1} Institute of Neuroscience and Medicine (INM-6) and Institute for Advanced Simulation (IAS-6) and JARA BRAIN Institute I, J\"ulich Research Centre, 52425 J\"ulich, Germany\\
{\bf 2} Department of Physics, Faculty 1, RWTH Aachen University, 52074 Aachen, Germany\\
\bigskip



* \href{mailto:t.kuehn@fz-juelich.de}{t.kuehn@fz-juelich.de}

\end{flushleft}

\section*{Abstract }

Population-wide oscillations are ubiquitously observed in mesoscopic
signals of cortical activity. In these network states a global oscillatory
cycle modulates the propensity of neurons to fire. Synchronous activation
of neurons has been hypothesized to be a separate channel of signal
processing information in the brain. A salient question is therefore
if and how oscillations interact with spike synchrony and in how far
these channels can be considered separate. Experiments indeed showed
that correlated spiking co-modulates with the static firing rate and
is also tightly locked to the phase of beta-oscillations.  While
the dependence of correlations on the mean rate is well understood
in feed-forward networks, it remains unclear why and by which mechanisms
correlations tightly lock to an oscillatory cycle. We here demonstrate
that such correlated activation of pairs of neurons is qualitatively
explained by periodically-driven random networks. We identify the
mechanisms by which covariances depend on a driving periodic stimulus.
Mean-field theory combined with linear response theory yields closed-form
expressions for the cyclostationary mean activities and pairwise zero-time-lag
covariances of binary recurrent random networks. Two distinct mechanisms
cause time-dependent covariances: the modulation of the susceptibility
of single neurons (via the external input and network feedback) and
the time-varying variances of single unit activities. For some parameters,
the effectively inhibitory recurrent feedback leads to resonant covariances
even if mean activities show non-resonant behavior. Our analytical
results open the question of time-modulated synchronous activity to
a quantitative analysis.

\section*{Author summary}

\ifthenelse{\boolean{isarxiv}}{}{\linenumbers}

In network theory, statistics are often considered to be stationary.
While this assumption can be justified by experimental insights to
some extent, it is often also made for reasons of simplicity. However,
the time-dependence of statistical measures do matter in many cases.
For example, time-dependent processes are examined for gene regulatory
networks or networks of traders at stock markets. Periodically changing
activity of remote brain areas is visible in the local field potential
(LFP) and its influence on the spiking activity is currently debated
in neuroscience. In experimental studies, however, it is often difficult
to determine time-dependent statistics due to a lack of sufficient
data representing the system at a certain time point. Theoretical
studies, in contrast, allow the assessment of the time dependent statistics
with arbitrary precision. We here extend the analysis of the correlation
structure of a homogeneously connected EI-network consisting of binary
model neurons to the case including a global sinusoidal input to the
network. We show that the time-dependence of the covariances - to
first order - can be explained analytically. We expose the mechanisms
that modulate covariances in time and show how they are shaped by
inhibitory recurrent network feedback and the low-pass characteristics
of neurons. These generic properties carry over to more realistic
neuron models.

\selectlanguage{english}%

\selectlanguage{american}%

\section*{Introduction}

To date it is unclear which channels the brain uses to represent and
process information. A rate-based view is argued for by the apparent
stochasticity of firing \citep{Softky93} and by the high sensitivity
of the network dynamics to single spikes \citep{London10_123}. In
an extreme view, correlated firing is a mere epiphenomenon of neurons
being connected. Indeed, a large body of literature has elucidated
how correlations relate to the connectivity structure \citep{Lindner05_061919,DeLaRocha07_802,Shea-Brown08,Renart10_587,Pernice11_e1002059,Trousdale12_e1002408,Pernice12_031916,Tetzlaff12_e1002596,Hu13_P03012,Helias13_023002,Helias14,Rosenbaum16_107}.
But the matter is further complicated by the observation that firing
rates and correlations tend to be co-modulated, as demonstrated experimentally
and explained theoretically \citep{DeLaRocha07_802,Shea-Brown08}.
If the brain employs correlated firing as a means to process or represent
information, this requires in particular that the appearance of correlated
events is modulated in a time-dependent manner. Indeed, such modulations
have been experimentally observed in relation to the expectation of
the animal to receive task-relevant information \citep{Riehle97_1950,Kilavik09_12653}
or in relation to attention \citep{Cohen09_1079}.

Oscillations are an extreme case of a time-dependent modulation of
the firing rate of cells. They are ubiquitously observed in diverse
brain areas and typically involve the concerted activation of populations
of neurons \citep{Buzsaki04_1926}. They can therefore conveniently
be studied in the local field potential (LFP) that represents a complementary
window to the spiking activity of individual neurons or small groups
thereof: It is composed of the superposition of the activity of hundreds
of thousands to millions of neurons \citep{Nunez06,Ray08_1529} and
forward modeling studies have confirmed \citep{Linden11_859} that
it is primarily driven by the synaptic inputs to the local network
\citep{Mitzdorf85_37,Viswanathan07_1308,Mazzoni2015}. As the LFP
is a quantity that can be measured relatively easily, this mesoscopic
signal is experimentally well documented. Its interpretation is, however,
still debated. For example, changes in the amplitude of one of the
components of the spectrum of the LFP have been attributed to changes
in behavior (cf. e.g. \citep{Scherberger05_347}).

A particular entanglement between rates and correlations is the correlated
firing of spikes in pairs of neurons in relation to the phase of an
ongoing oscillation. With the above interpretation of the LFP primarily
reflecting the input to the cells, it is not surprising that the mean
firing rate of neurons may modulate in relation to this cycle. The
recurrent network model indeed confirms this expectation, as shown
in \prettyref{fig:Time_dependent_correlations_illustrative}A. It
is, however, unclear if and by which mechanisms the covariance of
firing follows the oscillatory cycle. The simulation shown in \prettyref{fig:Time_dependent_correlations_illustrative}B
indeed exhibits a modulation of the covariance between the activities
of pairs of cells. Such modulations have also been observed in experiments:

Denker et al. \citep{Denker11_2681} have shown that the synchronous
activation of pairs of neurons within milliseconds preferentially
appears at a certain phase of the oscillatory component of the LFP
in the beta-range - in their words the spike-synchrony is ``phase-locked''
to the beta-range of the LFP. They explain their data by a conceptual
model, in which an increase in the local input, assumed to dominate
the LFP, leads to the activation of cell assemblies. The current
work investigates an alternative hypothesis: We ask if a periodically-driven
random network is sufficient to explain the time-dependent modulation
of covariances between the activities of pairs of cells or whether
additional structural features of the network are required to explain
this experimental observation.

 To investigate the mechanisms causing time-dependent covariances
in an analytically tractable case, we here present the simplest model
that we could come up with that captures the most important features:
A local network receiving periodically changing external input. The
randomly connected neurons receive sinusoidally modulated input, interpreted
as originating from other brain areas and mimicking the major source
of the experimentally observed LFP. While it is obvious that the mean
activity in a network follows an imposed periodic stimulation, it
is less so for covariances. In the following we will address the question
why they are modulated in time as well. Extending the analysis of
mean activities and covariances in the stationary state \citep{Helias14,Ginzburg94,VanVreeswijk98_1321},
we here expose the fundamental mechanisms that shape covariances in
periodically driven networks.

Our network model includes five fundamental properties of neuronal
dynamics: First, we assume that the state of low and irregular activity
in the network \citep{Softky93} is a consequence of its operation
in the balanced state \citep{Vreeswijk96,Amit-1997_373}, where negative
feedback dynamically stabilizes the activity. Second, we assume that
each neuron receives a large number of synaptic inputs \citep{Braitenberg91},
each of which only has a minor effect on the activation of the receiving
cell, so that total synaptic input currents are close to Gaussian.
Third, we assume the neurons are activated in a threshold-like manner
depending on their input. Fourth, we assume a characteristic time
scale $\tau$ that measures the duration of the influence a presynaptic
neuron has on its postsynaptic targets. Fifth, the output of the neuron
is dichotomous or binary, spike or no spike, rather than continuous.
As a consequence, the variance of the single unit activity is a direct
function of its mean.

We here show how each of the five above-mentioned fundamental properties
of neuronal networks shape and give rise to the mechanisms that cause
time-dependent covariances. The presented analytical expressions for
the linear response of covariances expose two different paths by which
a time-dependence arises: By the modulation of single-unit variances
and by the modulation of the linear gain resulting from the non-linearity
of the neurons. The interplay of negative recurrent feedback and direct
external drive can cause resonant behavior of covariances even if
mean activities are non-resonant. Qualitatively, these results explain
the modulation of synchrony in relation to oscillatory cycles that
are observed in experiments, but a tight locking of synchronous events
to a particular phase of the cycle is beyond the mechanisms found
in the here-studied models.

\section*{Results\label{sec:Results}}

\begin{figure}[h]
\begin{centering}
\includegraphics[scale=0.75]{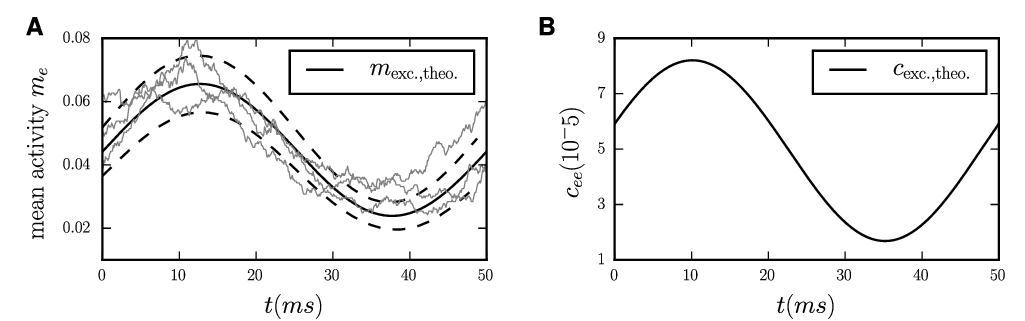}
\par\end{centering}
\caption{\label{fig:Time_dependent_correlations_illustrative}\foreignlanguage{american}{\textcolor{red}{
}\textbf{A} Time-varying mean activity of the excitatory population
$m_{E}(t)=N_{E}^{-1}\sum_{i\in E}\,n_{i}(t)$ in a balanced EI-network
(parameters given in \prettyref{tab:Biological_parameters}). Thin
gray lines are the outcomes of three independent simulations, the
solid black line indicates the mean activity predicted by the theory
(\prettyref{eq:m_self_consistent} and \prettyref{eq:mean_activity_Fourier}).
Dashed black lines indicate the range of expected fluctuations of
the population activity ($\pm$ one standard deviation): The square
of the fluctuation magnitude is given by the variance of the population
activity $\frac{a_{E}}{N_{E}}+c_{EE}$ (\prettyref{eq:Def_a} and
\prettyref{eq:Def_c_pairwise_pop}). \textbf{B} Population-averaged
cross covariance $c_{EE}=\frac{1}{N_{E}(N_{E}-1)}\,\sum_{i\protect\neq j\in\mathcal{E}}\,c_{ij}$.
}}
\end{figure}

To understand the locking of synchronous activity to an oscillatory
cycle, as observed experimentally, we here need to consider time-dependent
network states. We are in particular interested in the covariance
between two stochastic variables $x_{1}$ and $x_{2}$, which is defined
as $c(t)=\langle\delta x_{1}(t)\delta x_{2}(t)\rangle=\langle(x_{1}(t)-\langle x_{2}(t)\rangle)\,(x_{1}(t)-\langle x_{2}(t)\rangle\rangle$,
where $\left\langle ...\right\rangle $ denotes the average over realizations.
In words, the covariance in a time-dependent setting measures the
co-variability of a pair of signals with respect to their respective
mean. The mean value itself may depend on time.   Only if this quantity
can be determined with sufficient accuracy, time-dependent covariances
can be calculated correctly. This is the source of the technical problems
occurring in the context of a time-dependent covariance: It may be
hard to assess the covariance, much more its time-dependence, because
it is overshadowed by the time-varying mean. If a stochastic model
is given, however, disentangling the time dependence of different
cumulants, like mean and covariance, is possible. A theoretical study
to understand the prevalent mechanisms that cause time-dependent covariances
in a network model is therefore a necessary first step. In the following
we identify these mechanisms by which time-dependent covariances of
activities arise in oscillatory-driven recurrent networks. In \prettyref{fig:Time_dependent_correlations_illustrative}\textbf{A}
we show the population-averaged activity of the excitatory population
activity in a balanced EI-network together with the theoretical prediction
to be developed in the sequel: The fluctuations around the mean show
a wider spread close to the peak of the oscillation than at the trough.
Correspondingly, the covariance between pairs of neurons in panel
\textbf{B} has its peaks and troughs at points of high and low variability
of the population activity in \textbf{A}, respectively. 

\selectlanguage{english}%

\selectlanguage{american}%

\subsection*{Binary network model and its mean field equations\label{subsec:Results_formulas}}

\selectlanguage{english}%
\begin{figure}[h]
\begin{centering}
\includegraphics{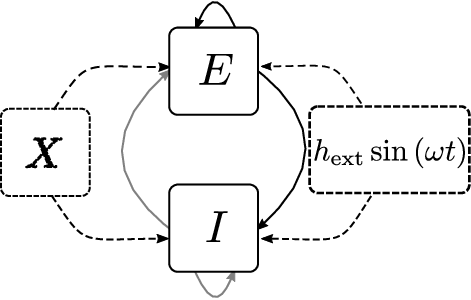}
\par\end{centering}
\centering{}\caption{\foreignlanguage{american}{\label{fig:Network_sketch}\textbf{Recurrent balanced network driven
by oscillatory input.} Recurrently connected excitatory ($E$) and
inhibitory ($I$) populations (Erd\H{o}s-R\'enyi random network
with connection probability $p$) receiving input from an external
($X$) excitatory population. Additionally, all neurons in the microcircuit
receive a sinusoidal signal of amplitude $\protect\hext$ and frequency
$\omega$, representing the oscillatory activity received from external
brain areas.}}
\end{figure}

\selectlanguage{american}%
To address our central question, whether a periodically-driven random
network explains the experimental observations of time-modulated pairwise
covariances, we consider a minimal model here. It consists of one
inhibitory ($I$) population and, in the latter part of the paper,
additionally one excitatory population ($E$) of binary model neurons
\citep{Renart10_587,Ginzburg94,Vreeswijk96,Vreeswijk98}. Neurons
within these populations are recurrently and randomly connected. All
neurons are driven by a global sinusoidal input mimicking the incoming
oscillatory activity that is visible in the LFP, illustrated in \prettyref{fig:Network_sketch}.
The local network may in addition receive input from an external excitatory
population ($X$), representing the surrounding of the local network.
The fluctuations imprinted by the external population, providing shared
inputs to pairs of cells, in addition drive the pairwise covariances
within the network \citep[c.f. especially the discussion]{Helias14}.
Therefore we need the external population $X$ to arrive at a realistic
setting that includes all sources of covariances. In the following,
we extend the analysis of cumulants in networks of binary neurons
presented in \citep{Renart10_587,Helias14,Ginzburg94,VanVreeswijk98_1321,Buice10_377}
to the time-dependent setting. This formal analysis allows us to obtain
analytical approximations for the experimentally observable quantities,
such as pairwise covariances, that expose the mechanisms shaping correlated
network activity.

Binary model neurons at each point in time are either inactive $n_{i}=0$
or active $n_{i}=1$. The time evolution of the network follows the
Glauber dynamics \citep{Glauber63_294}; the neurons are updated asynchronously.
At every infinitesimal time step $dt$, any neuron is chosen with
probability $\frac{dt}{\tau}$. After an update, neuron $i$ is in
the state $1$ with the probability $F_{i}(\boldsymbol{n})$ and in
the $0$-state with probability $1-F_{i}(\boldsymbol{n})$, where
the activation function $F$ is chosen to be
\begin{align}
F_{i}(\boldsymbol{n}) & =H\left(h_{i}-\theta_{i}\right)\nonumber \\
h_{i} & =\sum_{k=1}^{N}J_{ik}n_{k}+\hext\sin\left(\omega t\right)+\xi_{i}\label{eq:Def_gain_function}\\
H(x) & =\begin{cases}
1 & \text{if }x\ge0\\
0 & \text{if }x<0
\end{cases}.\nonumber 
\end{align}
We here introduced the connectivity matrix $J$ with the synaptic
weights $J_{ij}\in\mathbb{R}$ describing the influence of neuron
$j$ on neuron $i$. The weight $J_{ij}$ is negative for an inhibitory
neuron $j$ and positive for an excitatory neuron. Due to the synaptic
coupling the outcome of the update of neuron $i$ potentially depends
on the state $\boldsymbol{n}=(n_{1},\ldots,n_{N})$ of all other neurons
in the network. Compared to the equations in \citep[page 4]{Helias14},
we added an external sinusoidal input to the neurons representing
the influence of other cortical or subcortical areas and Gaussian
uncorrelated noise with vanishing mean $\left\langle \xi_{i}\right\rangle =0$
and covariance $\left\langle \xi_{i}\xi_{j}\right\rangle =\delta_{ij}\sigma_{\text{noise}}^{2}$.
The threshold $\theta_{i}$ depends on the neuron type and will be
chosen according to the desired mean activity.

We employ the neural simulation package NEST \citep{Gewaltig_07_11204,Nest280}
for simulations. Analytical results are obtained by mean-field theory
\citep{Renart10_587,Helias14,Ginzburg94,VanVreeswijk98_1321,Buice09_377,VanVreeswijk06_143}
and are described for completeness and consistency of notation in
the section \nameref{sec:Methods}. In the main text we only mention
the main steps and assumptions entering the approximations. The basic
idea is to describe the time evolution of the Markov system in terms
of its probability distribution $p\left(\boldsymbol{n},t\right)$.
Using the master equation \ref{eq:Master-equation} we obtain ordinary
differential equations (ODEs) for the moments of $p\left(\boldsymbol{n},t\right)$.
In particular we are interested in the population averaged mean activities
$m_{\alpha}$, variances $a_{\alpha}$, and covariances $c_{\alpha\beta}$
\begin{eqnarray}
m_{\alpha}\left(t\right) & := & \frac{1}{N_{\alpha}}\sum_{i\in\alpha}\left\langle n_{i}\left(t\right)\right\rangle \label{eq:Def_m_pop}\\
a_{\alpha}\left(t\right) & := & \frac{1}{N_{\alpha}}\sum_{i\in\alpha}\left\langle n_{i}\left(t\right)\right\rangle -\left\langle n_{i}\left(t\right)\right\rangle ^{2}\label{eq:Def_a}\\
c_{\alpha\beta}\left(t\right) & := & \frac{1}{N_{\alpha}N_{\beta}}\sum_{i\in\alpha,j\in\beta,i\neq j}\left\langle n_{i}\left(t\right)n_{j}\left(t\right)\right\rangle -\left\langle n_{i}\left(t\right)\right\rangle \left\langle n_{j}\left(t\right)\right\rangle ,\label{eq:Def_c_pairwise_pop}
\end{eqnarray}
which are defined as expectation values $\langle\rangle$ over realizations
of the network activity, where the stochastic update of the neurons
and the external noisy input presents the source of randomness in
the network. The dynamics couples moments of arbitrarily high order
\citep{Buice10_377}. To close this set of equations, we neglect cumulants
of order higher than two, which also approximates the input by a Gaussian
stochastic variable with cumulants that vanish for orders higher than
two \citep{Dahmen16_031024}. This simplification can be justified
by noticing that the number of neurons contributing to the input is
large and their activity is weakly correlated, which makes the central
limit theorem applicable. In a homogeneous random network, on expectation
there are $K_{\alpha\beta}=p_{\alpha\beta}N_{\beta}$ synapses from
population $\beta$ to a neuron in population $\alpha$. Here $p_{\alpha\beta}$
is the connection probability; the probability that there is a synapse
from any neuron in population $\beta$ to a particular neuron in population
$\alpha$ and $N_{\alpha}$ is the size of the population. Mean \prettyref{eq:Def_m_pop}
and covariance \prettyref{eq:Def_c_pairwise_pop} then follow the
coupled set of ordinary differential equations (ODEs, see section
\prettyref{subsec:Derivation_of_moment_ODEs} for derivation) 
\begin{eqnarray}
\tau\frac{d}{dt}m_{\alpha}\left(t\right) & = & -m_{\alpha}\left(t\right)+\varphi(\mu_{\alpha}(\boldsymbol{m}\left(t\right),\hext\sin\left(\omega t\right)),\sigma{}_{\alpha}(\boldsymbol{m}\left(t\right),c\left(t\right)))\label{eq:ODE_mean_population_averaged}\\
\tau\frac{d}{dt}c_{\alpha\beta}\left(t\right) & = & \Bigg\{-c_{\alpha\beta}\left(t\right)+\sum_{\gamma}\Bigg[S\left(\mu_{\alpha}\left(\boldsymbol{m}\left(t\right),\hext\sin\left(\omega t\right)\right),\sigma_{\alpha}\left(\boldsymbol{m}\left(t\right),c\left(t\right)\right)\right)\label{eq:ODE_correlation_population_averaged}\\
 &  & \times K_{\alpha\gamma}J_{\alpha\gamma}\,\Bigg(c_{\gamma\beta}\left(t\right)+\delta_{\gamma\beta}\frac{a_{\beta}\left(t\right)}{N_{\beta}}\Bigg)\Bigg]\Bigg\}+\left\{ \alpha\leftrightarrow\beta\right\} ,\nonumber 
\end{eqnarray}
where $\alpha\leftrightarrow\beta$ indicates the transposed term.
The Gaussian truncation employed here is parameterized by the mean
$\mu_{\alpha}$ and the variance $\sigma_{\alpha}^{2}$ of the summed
input to a neuron in population $\alpha$. These, in turn, are functions
of the mean activity and the covariance, given by \prettyref{eq:Def_mu_pop}
and \prettyref{eq:Def_sigma_pop}, respectively.

Here $\varphi$ is the expectation value of the activation function,
which is smooth, even though the activation function itself is a step
function, therefore not even continuous. The function $\varphi$ fulfills
$\lim_{\boldsymbol{m}\rightarrow0}\varphi=0$ and $\lim_{\boldsymbol{m}\rightarrow1}\varphi=1$
and monotonically increases. Its derivative $S$ with respect to
$\mu$ has a single maximum and is largest for the mean input $\mu$
within a region with size $\sigma$ around the threshold $\theta$.
$S$ measures the strength of the response to a slow input and is
therefore termed susceptibility. The definitions are given in \nameref{sec:Methods}
in \prettyref{eq:Def_phi_pop} and \prettyref{eq:Def_susceptibility_pop}.

The stationary solution (indicated by a bar) of the ODEs \prettyref{eq:ODE_mean_population_averaged}
and \prettyref{eq:ODE_correlation_population_averaged} can be found
by solving the equations\foreignlanguage{english}{
\begin{eqnarray}
\overline{\boldsymbol{m}} & = & \varphi\left(\overline{\boldsymbol{m}}\right)\label{eq:m_self_consistent}\\
2\overline{c} & = & SKJ\,\left(\overline{c}+\frac{\overline{a}}{N}\right)+\text{transposed}\label{eq:c_self_consistent}
\end{eqnarray}
}numerically and self-consistently, as it was done in \citep{Helias14,Ginzburg94,Buice10_377}.

 The full time-dependent solution of \prettyref{eq:ODE_mean_population_averaged}
and \prettyref{eq:ODE_correlation_population_averaged} can, of course,
be determined numerically without any further assumptions. Besides
the comparison with simulation results, this will give us a check
for the subsequently applied linear perturbation theory. The resulting
analytical results allow the identification of the major mechanisms
shaping the time-dependence of the first two cumulants. To this end,
we linearize the ODEs \prettyref{eq:ODE_mean_population_averaged}
and \prettyref{eq:ODE_correlation_population_averaged} around their
stationary solutions. We only keep the linear term of order $\hext$
of the deviation, justifying a Fourier ansatz for the solutions. For
the mean activities this results in $m_{\alpha}\left(t\right)=\overline{m}_{\alpha}+\delta m_{\alpha}\left(t\right)=\overline{m}_{\alpha}+M_{\alpha}^{1}e^{i\omega t}$
with 
\begin{equation}
M_{\alpha}^{1}=\sum_{\beta}U_{\alpha\beta}M_{\beta}^{1}=\sum_{\beta}U_{\alpha\beta}\frac{\hext\left(U^{-1}S\left(\overline{\boldsymbol{\mu}},\overline{\boldsymbol{\sigma}}\right)\right)_{\beta}\left(-i\tau\omega+1-\lambda_{\beta}\right)}{\left(\tau\omega\right)^{2}+\left(1-\lambda_{\beta}\right)^{2}}.\label{eq:mean_activity_Fourier}
\end{equation}

The time-dependence of $\sigma$ was neglected here, which can be
justified for large networks (\nameref{sec:Methods}, \prettyref{eq:Def_del_sig}
and \prettyref{eq:ODE_c_first_version}). The matrix $U$ represents
the basis change that transforms $\overline{W}_{\alpha\beta}:=S\left(\overline{\mu}_{\alpha},\overline{\sigma}_{\alpha}\right)\,K_{\alpha\beta}J_{\alpha\beta}$
into a diagonal matrix with $\lambda_{\alpha}$ the corresponding
eigenvalues. We see that, independent of the number of populations
or the detailed form of the connectivity matrix, the amplitude of
the time-dependent part of the mean activities has the shape of a
low-pass-filtered signal to first order in $\hext$. Therefore the
phase of $\delta\boldsymbol{m}$ lags behind the external drive and
its amplitude decreases asymptotically like $\frac{1}{\omega}$, as
can be seen in \prettyref{fig:Single_pop_frequency}A, B.

If we also separate the covariances into their stationary part and
a small deviation that is linear in the external drive, $c_{\alpha\beta}\left(t\right)=\overline{c}_{\alpha\beta}+\delta c_{\alpha\beta}\left(t\right)$,
expand $S\left(\mu_{\alpha}\left(t\right),\sigma_{\alpha}\left(t\right)\right)$
and $a\left(t\right)$ around their stationary values, keeping only
the terms of order $\hext$ and neglect contributions from the time-dependent
variation of the variance of the input $\sigma^{2}$ (see \nameref{sec:Methods},
especially \prettyref{eq:ODE_c_first_version} for a discussion of
this point), we get the ODE

\begin{eqnarray}
 &  & \tau\frac{d}{dt}\delta c\left(t\right)+2\delta c\left(t\right)-\overline{W}\delta c\left(t\right)-\left(\overline{W}\delta c\left(t\right)\right)^{T}\nonumber \\
 & = & \Bigg\{\underbrace{\overline{W}\,\mathrm{diag}\left(\frac{1-2\boldsymbol{\overline{m}}}{N}\right)\mathrm{diag}\left(\boldsymbol{\delta m}\left(t\right)\right)}_{\ainhom}\nonumber \\
 & + & \Big[\underbrace{\mathrm{diag}\left(K\circledast J\,\boldsymbol{\delta m}\left(t\right)\right)}_{\Sminhom}+\underbrace{\hext\sin\left(\omega t\right)}_{\Shinhom}\Big]\,\mathrm{diag}\left(\frac{\partial\boldsymbol{S}}{\partial\mu\left(t\right)}\right)\,K\circledast J\,\overline{c}^{\mathrm{total}}\Bigg\}\label{eq:ODE_compact_coloured}\\
 & + & \left\{ ...\right\} ^{T},\nonumber 
\end{eqnarray}
where we introduced the point-wise (Hadamard) product $\circledast$
of two matrices $A$ and $B$ \citep[see][ for a consistent notation of matrix operations]{cichocki09}
as $\left(A\circledast B\right)_{ij}:=A_{ij}B_{ij}$, defined the
matrix with the entries $\mathrm{diag}\left(\boldsymbol{x}\right)_{ij}:=\delta_{ij}x_{i}$
for the vector $\boldsymbol{x}=\left(x_{1},..,x_{n}\right)$ and set
$\overline{c}^{\mathrm{total}}:=\overline{c}+\mathrm{diag}\left(\frac{\overline{a}}{N}\right)$
to bring our main equation into a compact form. 

We can now answer the question posed in the beginning: Why does a
global periodic drive influence the cross covariances in the network
at all and does not just make the mean activities oscillate? First,
the variances are modulated with time, simply because they are determined
via \prettyref{eq:Def_a} by the modulated mean activities. A neuron
$i$ with modulated autocorrelation $a_{i}\left(t\right)$ projects
via $J_{ji}$ to another neuron $j$ and therefore shapes the pairwise
correlation $c_{ji}(t)$ in a time-dependent way. We call this effect
the ``modulated-autocovariances-drive'', indicated by the curly
brace in the second line of \eqref{eq:ODE_compact_coloured}. Its
form in index notation is $[\overline{W}\,\mathrm{diag}\left((1-2\boldsymbol{\overline{m}})/N\right)\mathrm{diag}\left(\boldsymbol{\delta m}\left(t\right)\right)]_{\alpha\beta}=\overline{W}_{\alpha\beta}\,(1-2\overline{m}_{\beta})/N_{\beta}\,\delta m_{\beta}(t)$.
This is the low-pass-filtered input.

The other contributions are a bit more subtle and less obvious, as
they are absent in networks with a linear activation function. The
derivative of the expectation value of the activation function, the
susceptibility, contributes linearly to the ODE of the covariances.
As the threshold-like activation function gives rise to a nonlinear
dependence of $\varphi$ on the mean input $\mu$, the susceptibility
$S=\varphi^{\prime}$ is not constant, but depends on the instantaneous
mean input. The latter changes as a function of time by the direct
external drive and by the recurrent feedback of the oscillating mean
activity, indicated by the terms denoted by the curly braces in the
third line of \eqref{eq:ODE_compact_coloured}. Together, we call
these two term the ``susceptibility terms''. Both terms are of the
same form
\begin{align}
[\mathrm{diag}\left(\mathbf{\boldsymbol{\delta\mu}}(t)\right)\mathrm{diag}\left(\frac{\partial\boldsymbol{S}}{\partial\mu\left(t\right)}\right)\,K\circledast J\,\overline{c}^{\mathrm{total}}]_{\alpha\beta} & =\delta\mu_{\alpha}(t)\,\frac{\partial S_{\alpha}}{\partial\mu_{\alpha}}\,\sum_{\gamma}\,K_{\alpha\gamma}J_{\alpha\gamma}\,(\overline{c}_{\gamma\beta}+\delta_{\gamma\beta}\frac{\overline{a}_{\beta}}{N_{\beta}}),\label{eq:susceptibility_drive explicit}
\end{align}
but with different $\delta\mu_{\alpha}$. This form shows how the
time-dependent modulation of the mean input $\delta\mu_{\alpha}$,
by the second derivative of the gain function $\frac{\partial S_{\alpha}}{\partial\mu_{\alpha}}=\varphi^{\prime\prime}$,
influences the transmission of covariances. The sum following $\frac{\partial S_{\alpha}}{\partial\mu_{\alpha}}$
is identical to the one in the static case \prettyref{eq:c_self_consistent}.
For the ``recurrent drive'', the time-dependent input is given by
$\delta\mu_{\alpha}(t)=\sum_{\beta}K_{\alpha\beta}J_{\alpha\beta}\delta m_{\beta}(t)$,
which is a superposition of the time-dependent activities that project
to population $\alpha$ and is therefore low-pass-filtered, too. The
term due to ``direct drive'' is $\delta\mu_{\alpha}(t)=\hext\sin\left(\omega t\right)$.

We solve \prettyref{eq:ODE_compact_coloured} by transforming into
the eigensystem of $\overline{W}$ and inserting a Fourier ansatz,
$\delta c_{\alpha\beta}\left(t\right)=C_{\alpha\beta}^{1}e^{i\omega t}$.
The solution consists of a low-pass filtered part coming from the
direct drive and two parts that are low-pass filtered twice, coming
from the recurrent drive and the modulated-autocovariances-drive.
For a detailed derivation, consult the section \nameref{subsec:Methods_correlations}.

We have calculated higher Fourier modes of the simulated network activity
and of the numerical solution of the mean-field equations to check
if they are small enough to be neglected, so that the response is
dominated by the linear part. Of course, it would be possible to derive
analytical expressions for those as well. However, we will see that
the linear order and the corresponding first harmonic qualitatively
and for remarkably large perturbations even quantitatively gives the
right predictions. The limits of this approximation are analyzed in
\prettyref{fig:Dependence_on_h}. We will therefore constrain our
analysis to controlling the higher harmonics through the numerical
solution.

In the following we will study three different models of balanced
neuronal networks to expose the different mechanisms in their respective
simplest setting.

\subsubsection*{Single population}

As a first example, we quantitatively study the particular case of
a single population, which has to be inhibitory to ensure stable stationary
activity. Let us look at the behavior of the different contributions
in \prettyref{eq:ODE_compact_coloured} to the modulated covariance
and their mutual relation. Written explicitly, the terms driving the
time variation of the covariance are
\begin{eqnarray}
 &  & \Big(\overbrace{\underbrace{KJ\,\delta m\left(t\right)}_{\Smt\ \propto\frac{1}{\omega}\text{ for big }\omega}+\underbrace{\hext\sin\left(\omega t\right)}_{\Sht\text{ does not scale with }\omega}}^{\Sinhom;\text{partly cancel}}\Big)\,\frac{\partial\boldsymbol{S}}{\partial\mu}KJ\underbrace{\left(\overline{c}+\frac{\overline{a}}{N}\right)}_{\text{partly cancel}}\label{eq:decomposition_single_pop}\\
 & + & \underbrace{\overline{W}\,\left(1-2\overline{m}\right)\,\frac{\delta\boldsymbol{m}\left(t\right)}{N}}_{\text{\ensuremath{\at}\ \ensuremath{\propto\frac{1}{\omega}\text{ for big }\omega}}}.\nonumber 
\end{eqnarray}
 With respect to their dependence on the number of synaptic connections
$\left|K\right|$, the sum of the two susceptibility terms is of the
same order of magnitude as the modulated-autocovariances-drive (cf.
\ref{eq:} in the section \nameref{sec:Methods}), therefore their
interplay determines the shape of the solution of \prettyref{eq:ODE_compact_coloured}
and we cannot neglect either term in favor of the other.

To analyze the contributions to $\delta c$, it is reasonable to first
focus on the quasi-static case $\omega\to0$, because its analysis
is simplest and, due to the continuity of the observed quantities,
it carries over to the case of biologically relevant small frequencies
up to the $\beta$-range. For $\omega\to0$, the solution $\delta c$
in \prettyref{eq:ODE_compact_coloured} has the same sign as the sum
of the inhomogeneities, because it is given by a multiplication with
$0.5\left(1-W\right)^{-1}$, where $W<0$. The main information -
especially about the sign - is therefore already included in these
inhomogeneities, that we termed ``recurrent drive'' and ``direct
drive'' (the susceptibility terms) and ``modulated-autocovariances-drive''
in the previous section. The modulation of the covariance $\delta c\left(t\right)$
then results by low pass filtering their sum. Individually they yield
the $S_{m}$-term and $S_{h}$-term (together the $S$-terms) and
the $a$-term, respectively. 

\begin{figure}[h]
\includegraphics[scale=0.75]{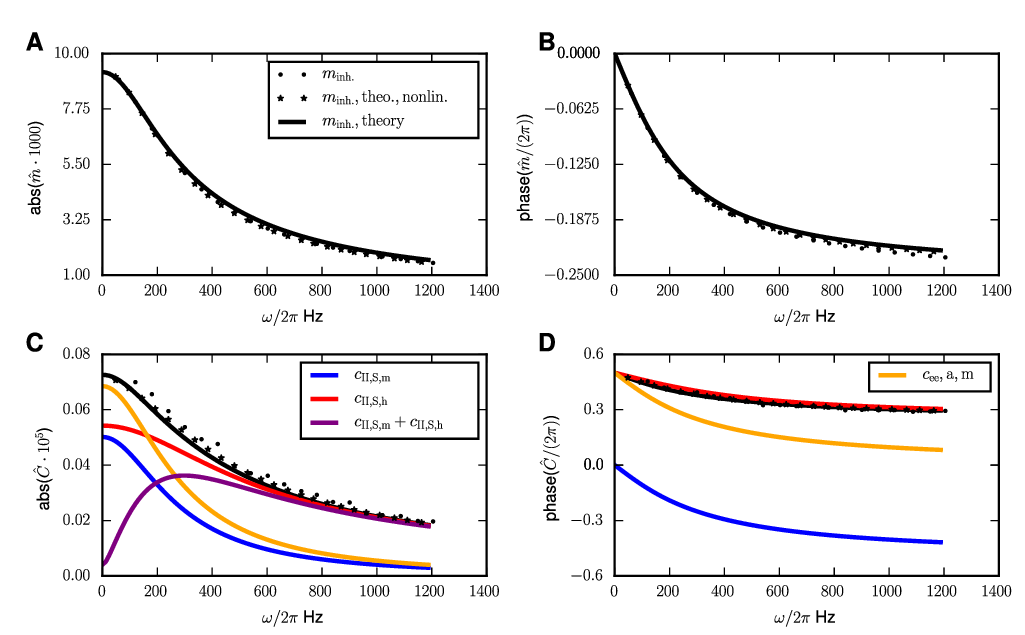}

\caption{\label{fig:Single_pop_frequency}\foreignlanguage{american}{\textbf{Periodically
driven single population network.} Dependence of the modulations of
the mean activity and covariances on the driving frequency $\omega$.
\textbf{A} Amplitude of modulation of mean activity. \textbf{B} Phase
of modulation of mean activity relative to the external drive. \textbf{C}
Amplitude of modulation of covariances. \textbf{D} Phase of modulation
of covariance relative to the external drive. In all panels, the
analytical predictions (\prettyref{eq:mean_activity_Fourier} and
\prettyref{eq:Correlation_Fourier}) are shown as solid black curves.
The black curve is the complete solution. The different contributions
to the time-dependent covariances, identified in \prettyref{eq:decomposition_single_pop},
are shown separately: The $S_{h}$-term in red, the $S_{m}$-term
in blue, their sum in purple, and the $a$-term in orange. Numerical
solutions of the full mean-field equations (\prettyref{eq:ODE_mean_population_averaged}
and \prettyref{eq:ODE_correlation_population_averaged}) are shown
as stars and simulation results by dots (only indicated in the legend
of A). The numerical results are obtained by using the integrate.ode-method
from the python-package scipy \citep{scipy01} with the option ``lsoda'',
meaning that either implicit Adams- or backward differentiation-algorithms
(depending on the given problem) are used. Network parameters: Number
of neurons $N_{I}=5000,$ connection probability $p_{II}=0.1$, coupling
strength $J_{II}=-1$, mean activity $m_{I}\approx0.3$, and $\sigma_{\text{noise}}=\sigma_{\text{system}}:=\sqrt{J_{II}^{2}p_{II}N_{I}m_{I}\left(1-m_{I}\right)}\approx10.2$.}
}
\end{figure}

In a general balanced network, the deviation of the mean activity
from the stationary solution $\delta\boldsymbol{m}\left(t\right)$
is in phase with the perturbation for $\omega\approx0$ and lags behind
it for larger $\omega$ due to the ``forgetfulness'' of the network
caused by the leak term in the ODE. At low frequencies, the recurrent
drive $\propto K\circledast J\,\boldsymbol{\delta m}\left(t\right)$
therefore partly cancels the direct drive $\propto\hext\sin\left(\omega t\right)$.
This is because the rate response $\delta m$ is in phase, and the
feedback $KJ<0$ in the network is negative. The cancellation becomes
less efficient at larger frequencies, because the recurrent drive
asymptotically decays like $\omega^{-1}$ and is phase-shifted; the
mean activity is low-pass-filtered \eqref{eq:mean_activity_Fourier}.
The direct drive, in contrast, does not depend on the driving frequency
$\omega$. Therefore, the $S_{m}$-term is low-pass-filtered twice
and the $S_{h}$-term term only once, therefore their sum has a peak
at an intermediate frequency, as visible in \prettyref{fig:Single_pop_frequency}C,
purple curve. Note that this cancellation generally appears in the
balanced state, because the network feedback is always effectively
inhibitory. Furthermore, the modulated-autocovariances-drive only
vanishes for $\overline{m}_{\alpha}=\frac{1}{2}$; for realistic activity
$\overline{m}_{\alpha}\ll\frac{1}{2}$ it is in anti-phase with $\delta m\left(t\right)$,
because it is defined including $W<0$, which flips the phase by $\pi$.

 Average covariances in inhibitory networks are negative \citep{Helias14}.
As a consequence, in the setting of a single inhibitory population
there is a second kind of cancellation: The two terms $\bar{c}$ and
$N^{-1}\bar{a}$ in the prefactor $\overline{c}+N^{-1}\overline{a}$
of the susceptibility terms in \prettyref{eq:decomposition_single_pop}
partly cancel; their sum in fact vanishes in the large $N$ limit
\citep[cf. ][eq. (32) and their Fig 5]{Helias14}. This leads to the
dominance of the $a$-term, shown in \prettyref{fig:Single_pop_frequency}D
(orange curve). The maximum in the $S$-terms is therefore overshadowed
by the $a$-term, which asymptotically also shows a second order low
pass characteristics with $\propto\omega^{-2}$. So in the purely
inhibitory network the peak is not visible in the sum of all contributions
to $\hat{C}\left(\omega\right)$.  

\selectlanguage{english}%

\selectlanguage{american}%
In summary, the model of a single population in the balanced state
exposes several generic features of time-dependent mean activities
and covariances: Mean activities and the direct drive contribution
to covariances follow the external modulation with first order low
pass characteristics. The $S_{m}$-term and the $a$-term of the covariances,
being mediated by the mean activity, consequently expose a second
order low pass filtering. The direct drive and the recurrent drive
(the susceptibility terms) to large extent cancel at low frequencies,
but not at high ones. Due to their overall decay in amplitude with
increasing frequency, an intermediate maximum arises in their sum.
In the single population model this peak is typically overshadowed
by the $a$-term. This is because of the suppression of population
fluctuations by negative feedback in the stationary state \citep{Tetzlaff12_e1002596},
which causes a small population variance $N^{-1}\overline{a}+\overline{c}$
and the latter term controls the amplitude of the susceptibility terms.

\subsubsection*{Two homogeneously connected populations}

A slightly more realistic, but still simple setup is an EI-network
with the same input for the inhibitory and the excitatory neurons,
as studied before, in \citep[ parameters, except $m_X$ as in fig. 6 there]{Helias14}.
This network is also inhibition-dominated, therefore we observe qualitatively
the same competition of the two $S$-terms leading to the existence
of a maximum in the $\omega$-dependence of $\left|C_{1}\right|$.
In contrast to the single population case, in the E-I network the
peak may be visible. This is because - in contrast to the single population
case - covariances in this setup may also be positive, preventing
the cancellation with the variances in the term $\overline{c}+\frac{\overline{a}}{N}$
that drives the $S$-terms. The latter contribution may therefore
dominate over the $a$-term at small $\omega$. Its dominance increases
the larger the covariances are, which for example arises when increasing
the external drive or by lowering the noise level at the input to
the neurons. The ``resonance'' effect itself increases for weaker
the excitatory synapses. 

\begin{figure}[h]
\centering{}\includegraphics[scale=0.75]{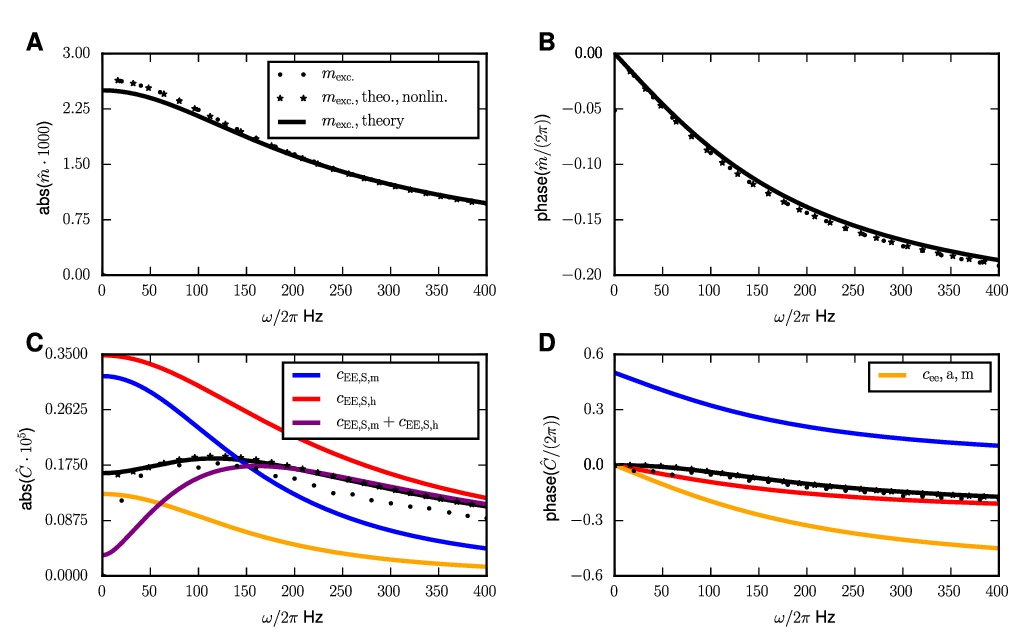}\caption{\label{fig:Homogeneous_frequency}\foreignlanguage{american}{\textbf{Periodically
driven E-I network.} \textbf{A} Amplitude of modulation of the mean
activity deviating from the stationary value for the excitatory population.
\textbf{B} Phase of the modulation of the mean activity. \textbf{C}
Different contributions to the amplitude of the covariance between
pairs of excitatory cells in dependence of the frequency $\omega$
of the external drive. \textbf{D} Phase of covariances relative to
the driving signal. Analytical theory (\prettyref{eq:mean_activity_Fourier},
\prettyref{eq:Correlation_Fourier}) shown by solid black curves,
numerical solutions of the full mean field equations (\prettyref{eq:ODE_mean_population_averaged}
and \prettyref{eq:ODE_correlation_population_averaged}) (stars) and
simulation results (dots, only indicated in the legend of A). Same
color code as in \textbf{C}. In \textbf{C} and \textbf{D}, the contributions
to the variation of covariances are shown separately: The $S_{h}$-term
in red, the $S_{m}$-term in blue, their sum in purple and the $a$-term
in yellow. The legend for \textbf{C }and \textbf{D} is split over
both panels. Numerical solutions obtained by the same methods as in
\prettyref{fig:Single_pop_frequency}. Parameters: $N_{E}=N_{I}=N_{X}=8192,\ p_{E}=p_{I}=p_{X}=p=0.2,\ m_{E}=m_{I}\approx0.11,\ m_{X}=0.25$,
identical to \citep[e.g. fig. 6]{Helias14}.}}
\end{figure}

\prettyref{fig:Homogeneous_frequency}C, indeed shows a peak of the
response of the covariances at a frequency of about $120\Hz$. We
here focus on the covariances between excitatory neurons, because
their activities are recorded most often and also cell assemblies
are normally assumed to consist of excitatory neurons.

\selectlanguage{english}%

\selectlanguage{american}%

\subsubsection*{Two populations with inhomogeneous connections\label{subsec:Two-populations-with}}

\begin{figure}[H]
\begin{centering}
\includegraphics[scale=0.75]{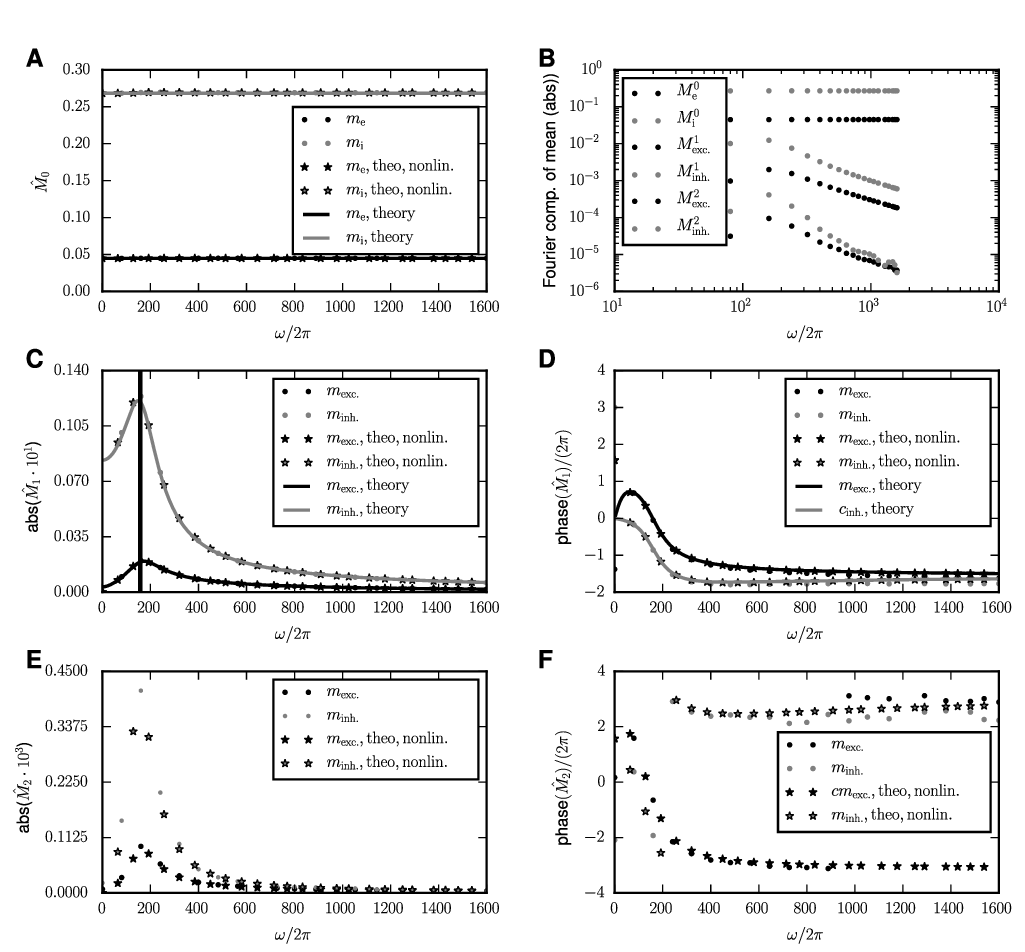}
\par\end{centering}
\caption{\foreignlanguage{american}{\textbf{Driven E-I network with biologically inspired parameters:
Mean activity.} From the first to the third row, the zeroth to second
Fourier mode of the mean activity is shown.\textbf{ A} Constant part
of mean activity (zeroth order). \textbf{B} First three Fourier-modes
of the mean activities on a loglog-scale.\textbf{ C} Amplitude of
first mode of the mean activity. \textbf{D} Phase of first mode relative
to driving signal. \textbf{E} and \textbf{F} are structured analogous
to \textbf{C} and \textbf{D} for the second Fourier modes. Solid curves
indicate the linear theory (\prettyref{eq:mean_activity_Fourier}),
stars numerical integration of the full mean field equations (\prettyref{eq:ODE_mean_population_averaged},
\prettyref{eq:ODE_correlation_population_averaged}) and dots the
simulation results of the full network. Black symbols indicate the
activity of excitatory, gray symbols of inhibitory neurons. Numerical
results obtained by the same methods as in \prettyref{fig:Single_pop_frequency}.
Noise amplitudes $\sigma_{\mathrm{noise},E}=\sigma_{\mathrm{noise},I}=10$,
$\sigma_{\mathrm{network},E}=2.8$, $\sigma_{\mathrm{network},E}=4.6$,
other parameters of the network model given in \prettyref{tab:Biological_parameters}.\label{fig:Biological_meanact_frequency}}}
\end{figure}

The example of homogeneous connectivity helps to explain the fundamental
mechanisms that shape the covariances; it is, however, certainly not
very realistic. Furthermore, in the case of synaptic weights being
different for individual receiving populations, the linearized connectivity
$W$ can have a pair of complex eigenvalues, which is qualitatively
different to the setup described before. To check if the theory also
works for parameters satisfying biological constraints, we choose
the connectivity and activity levels in accordance to experimental
studies. Apart from the results from \citep{Denker11_2681}, the parameters
were measured in the layer 2/3 in the barrel cortex of mice. We select
this layer, because it is the assumed location of cell assemblies
\citep{Yoshimura05_1552}, allowing us to relate our results to the
original hypothesis of excess synchrony by activation of assemblies
\citep{Denker11_2681}, a feature that could be considered in future
studies. The connection probabilities are taken from \citep{Avermann12},
the fractions of excitatory and inhibitory neurons from \citep{Lefort2009_301}
and the membrane time constant is extracted from \citep[supplementary material]{Gentet10_422}.
We adjust the neurons' thresholds such that the stationarity condition
$\boldsymbol{\varphi}\left(\boldsymbol{\overline{m}}\right)=\overline{\boldsymbol{m}}$
is fulfilled for $m_{\alpha}=\tau\nu_{\alpha}$, where $\alpha\in\left\{ \text{exc., inh.}\right\} $,
$\nu_{\alpha}$ is the firing rate of the respective population and
$\tau$ is the neuronal time constant. Note that the mapping $m=\tau\nu$
implies a slightly different notion of a ``spike'' of a binary neuron
than previously used \citep{VanVreeswijk98_1321}. The two conventions
agree in the limit of vanishing firing rates (cf. appendix, \nameref{subsec:Spike_defs}).
The firing rate of $18\Hz$ given in \citep{Denker11_2681} presumably
reflects the activity of excitatory neurons (private communication).
To obtain the firing rate of the inhibitory neurons $\nu_{\mathrm{inh.}}$,
we scale the measurement from \citep{Denker11_2681} by the ratio
$\nu_{\mathrm{\mathrm{inh}}}/\nu_{\mathrm{\mathrm{exc.}}}$ from \citep{Lefort2009_301}.
All parameters are summarized in \prettyref{tab:Biological_parameters}.
The effective connectivity $W$ of this system has two conjugate complex
eigenvalues. Therefore, there exists a resonance frequency also for
the mean activity, shown in \prettyref{fig:Biological_meanact_frequency}C.

\begin{table}[h]
\centering{}%
\begin{tabular}{|c|c|c|c|c|c|c||c|}
\hline 
 &  & exc. & inh. &  $\nu\,\left(\mathrm{Hz}\right)$ & mean act. & \#neurons & \tabularnewline
\hline 
\hline 
\multirow{2}{*}{exc.} & connection prob. & 0.168 & 0.5 & \multirow{2}{*}{18} & \multirow{2}{*}{0.045} & \multirow{2}{*}{1691} & \multirow{2}{*}{$\tau=2.5\,\mathrm{ms}$}\tabularnewline
\cline{2-4} 
 & synaptic weight & 0.37 & -0.52 &  &  &  & \tabularnewline
\hline 
\multirow{2}{*}{inh.} & connection prob. & 0.327 & 0.36 & \multirow{2}{*}{108} & \multirow{2}{*}{0.27} & \multirow{2}{*}{230} & \multirow{2}{*}{$m_{\mathrm{ext}}=0.1$}\tabularnewline
\cline{2-4} 
 & synaptic weight & 0.82 & -0.54 &  &  &  & \tabularnewline
\hline 
\end{tabular}\caption{\label{tab:Biological_parameters}Parameters for the biologically
inspired network model used in \prettyref{fig:Biological_meanact_frequency},
\prettyref{fig:Biological_c_ee_three_harm} and  \prettyref{fig:C_ii_three_harm}
and \prettyref{fig:C_ei_three_harm}.}
\end{table}

\begin{figure}[H]
\centering{}\includegraphics[scale=0.75]{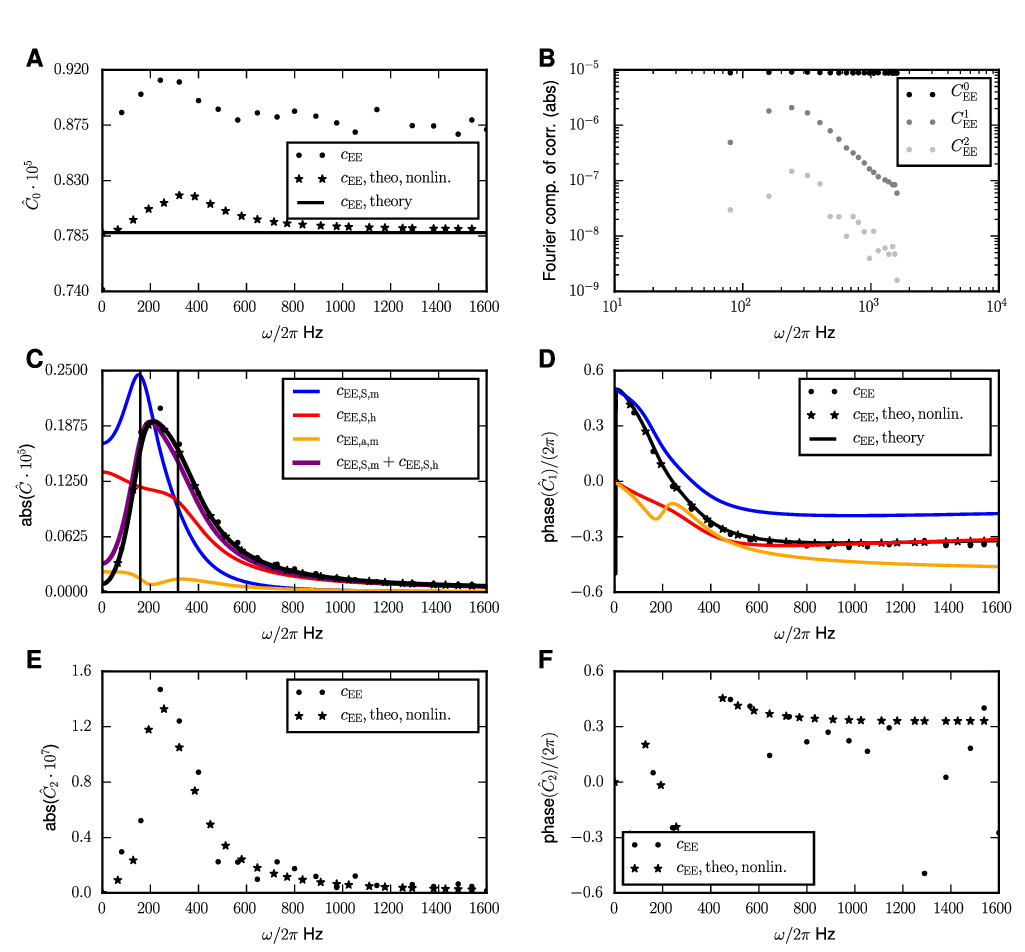}\caption{\foreignlanguage{american}{\textbf{Driven E-I network with biologically inspired parameters:
EE-Covariance.} Response of the covariance to a perturbation with
frequency $\omega$ in the Fourier space.\textbf{ A} Zeroth Fourier
mode (time independent part) of the covariance.\textbf{ B} Absolut
value of the first three Fourier components of the $c_{ee}$-covariances
on a loglog-scale.\textbf{ C} Absolute value of the first order of
the time-dependent part of the covariance. \textbf{D} Phase angle
in relation to the driving signal. \textbf{E} and \textbf{F} are analogous
to \textbf{C} and \textbf{D} for the second Fourier modes. Solid lines
indicate the linear theory \prettyref{eq:Correlation_Fourier}, stars
the results of the numerical solution of the full mean-field theory
\prettyref{eq:ODE_mean_population_averaged} and \prettyref{eq:ODE_correlation_population_averaged}
and dots the direct simulation of the full network\foreignlanguage{english}{.
}Numerical results obtained by the same methods as in \prettyref{fig:Single_pop_frequency}.\foreignlanguage{english}{
}Parameters of the network model as in \prettyref{fig:Biological_meanact_frequency}.\foreignlanguage{english}{\label{fig:Biological_c_ee_three_harm}}}}
\end{figure}

In the two upper panels of \prettyref{fig:Biological_meanact_frequency}
and \prettyref{fig:Biological_c_ee_three_harm}, we compare the stationary
values for the mean activity \prettyref{eq:m_self_consistent} and
the covariances \prettyref{eq:c_self_consistent} with the respective
time averaged results of the simulation and with the numerical solution
of the full mean-field equations. The stationary statistics have been
investigated before for other parameters in finite networks \citep{Helias14}
and in the limit $N\to\infty$ \citep{Renart10_587}. The second harmonics
extracted from the simulations and the numerical solution of the full
mean-field equations show good agreement and are overall small compared
to the zeroth and first harmonics, justifying the truncation of the
Fourier series in the analytical theory after the first term.

The first harmonics of the mean activity (see \prettyref{fig:Biological_meanact_frequency})
and covariances (see \prettyref{fig:Biological_c_ee_three_harm})
predicted by the linear response theory agree well with simulations
and the numerical solution. This is not necessarily clear a priori
because the perturbation in the input to every neuron is of the order
$\mathcal{O}\left(\frac{\sigma}{10}\right)$, where $\sigma$ is the
input noise level of the unperturbed system. However, linear response
theory works surprisingly well, even for the covariances caused by
a perturbation leading to a response of the same order of magnitude
as the stationary value. Increasing the perturbation strength $\hext$
further ultimately leads to a breakdown of the linear perturbation
theory, visible in the growing absolute values of the second Fourier
modes of mean activities and covariances (\prettyref{fig:Dependence_on_h}).
The maximal modulation in the firing rates amounts to $\approx0.8\Hz$
for the excitatory and $4.9\Hz$ for the inhibitory neurons.

 In this biologically inspired setting, it is also interesting to
apply the Unitary Event (UE) analysis to our data, as it was done
for experimental data in \citep{Denker11_2681}. Because this is a
bit aside the scope of this paper, we present this part in the appendix,
section \nameref{subsec:Application-of-the}.  

The connectivity matrix has complex eigenvalues $\lambda_{1}=\lambda_{2}^{*}$,
so we observe a resonance of the mean activities at the frequencies
\[
f_{\mathrm{res},\mathrm{mean}}=\frac{\Im\left(\lambda_{1}\right)}{\tau2\pi},
\]
indicated by a vertical line in \prettyref{fig:Biological_meanact_frequency}\textbf{C}.
The components of $\delta m$ are composed of different modes, therefore
their maximum does not appear exactly at $f_{\mathrm{res},\mathrm{mean}}$.
The covariances are shaped by more modes: In general, the covariance
matrix for a three-dimensional quantity has $6$ independent components.
In our case, $c_{XX}$ is always $0$, which is a consequence of the
missing feedback to $X$. Now, the evolution of every mode of $\widetilde{\delta c}$
is given by the sum of two eigenvalues of $1-W$, i.e. $2-\lambda$,
$2-\lambda^{*}$, $2-2\lambda$, $2-2\lambda^{*}$ and $2-\left(\lambda^{*}+\lambda\right)$.
The missing mode is the ``trivial'' one owing to the vanishing eigenvalue
of $W$. So the behavior of the ``kernel'' of the ODE for $\delta c$
is given by the resonances at $\frac{\left|\Im\left(\lambda\right)\right|}{\tau2\pi}$
and $2\cdot\frac{\left|\Im\left(\lambda\right)\right|}{\tau2\pi}$.
In addition, the inhomogeneity of the ODE (\ref{eq:ODE_compact_coloured})
(its right hand side) is already resonant at $\frac{\left|\Im\left(\lambda\right)\right|}{\tau2\pi}$.
All these modes are mixed with different strength in the different
modes of $\delta c$, giving rise to a maximum of $\left|C_{1}\right|$
somewhere in the vicinity of $f_{\mathrm{res},\mathrm{mean}}$ and
$2f_{\mathrm{res},\mathrm{mean}}$. In all cases the ``resonances''
are damped, therefore, a resonance catastrophe, induced by $\delta m$
oscillating with the resonance frequency of $\delta c$, cannot occur.
We also notice that all resonances are the stronger, the closer $\Re\left(\lambda\right)$
is to $1$, the critical point, which makes sense intuitively: The
damping comes from the overall inhibitory feedback; at the critical
point the leak term is exactly compensated by positive feedback of
identical magnitude. It is worth noticing that the effect of the
partial cancellation of the $S$-terms, which can be read off from
\prettyref{eq:ODE_compact_coloured} and is described in the previous
subsections for small $\omega$, is still valid. The functional form
of $\left|C_{1}\left(\omega\right)\right|$, however, is now mainly
determined by the resonances due to the complex eigenvalues of $W$.

The $\omega$-dependencies of the $c_{\mathrm{II}}$- and the $c_{\mathrm{EI}}$-
covariances shown in the appendix are qualitatively similar (\prettyref{fig:C_ii_three_harm}
and \prettyref{fig:C_ei_three_harm}). The stationary covariance is
well predicted by the theory \citep{Helias14}, which is confirmed
here. 

\selectlanguage{english}%

\begin{figure}[h]
\begin{centering}
\includegraphics[scale=0.75]{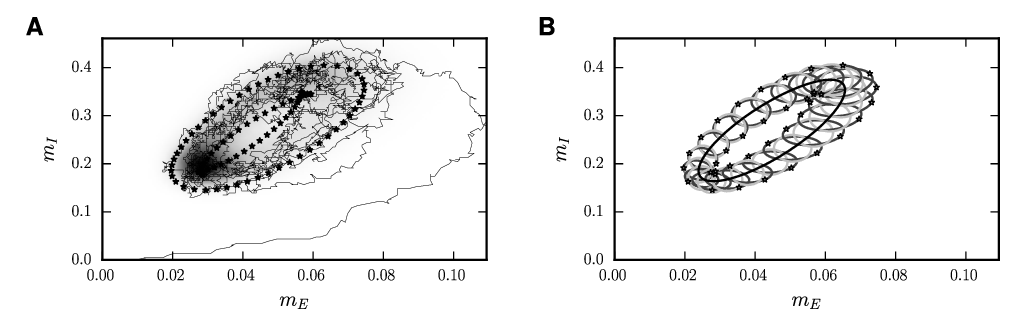}
\par\end{centering}
\caption{\foreignlanguage{american}{\textbf{Distribution of population-averaged activity of periodically
driven E-I network.} \textbf{A} Empirical density of population activity
of the E-I network. Gray shading indicates time-averaged occupation
of states. The thin mid gray curve is a sample of the binary dynamics
of $10$ periods after the start of the simulation. The black dots
indicate the $\sigma$-region predicted by the linear theory as described
by \prettyref{eq:sigma_region} in the main text. \textbf{B} Limit
cycle of the linear theory (black ellipse), together with error ellipses
stemming from the sum of covariances and variances (dark gray, slightly
tilted) and representing solely variances (light gray). The stars
are at the same places as in A. Parameters are given in \prettyref{tab:Biological_parameters},
only the perturbation strength is increased to $\protect\hext=6$
(noise level around $\sigma_{E}\simeq14$, $\sigma_{I}\simeq23$)
for reasons of readability (for this value the simulated results already
show deviations from the linear approximation as shown in  \prettyref{fig:Dependence_on_h}).
The perturbing frequency is chosen to be $f=80\protect\Hz$. \foreignlanguage{english}{\label{fig:Rate_dynamics-1}}}}
\end{figure}

\foreignlanguage{american}{\prettyref{fig:Rate_dynamics-1} illustratively
summarizes the results of this section. In panel A, the probability
of the binary system to be in a certain activity state $\left(\mI,\mE\right)^{\mathrm{T}}$
is indicated by different gray shades, the darker, the higher the
probability to find it in the respective area. On top, the area including
the most probable network states, as predicted by the linear theory,
is indicated by black dots. Its construction is depicted in panel
B: We draw the limit cycle (black) formed by the points $\left(\left\langle \mI\left(t\right)\right\rangle ,\left\langle \mE\left(t\right)\right\rangle \right)^{\mathrm{T}}$
as a parametric plot with time as parameter. Then, we define the points
on the error ellipse $\left(\mI,\mE\right)^{\mathrm{T}}$ as follows
\begin{equation}
\delta\boldsymbol{m}\left(t\right)^{\mathrm{T}}\left(c^{\mathrm{pop}}\left(t\right)\right)^{-1}\delta\boldsymbol{m}\left(t\right)=1,\label{eq:sigma_region}
\end{equation}
where $\delta\boldsymbol{m}^{T}:=\left(\mI,\mE,0\right)^{\mathrm{T}}-\left(\left\langle \mI\right\rangle ,\left\langle \mE\right\rangle ,0\right)^{\mathrm{T}}$
and 
\[
c^{\mathrm{pop}}\left(t\right)=\left(\begin{array}{ccc}
c_{\mathrm{EE}}^{\mathrm{pop}}\left(t\right) & c_{\mathrm{EI}}^{\mathrm{pop}}\left(t\right) & c_{\mathrm{EX}}^{\mathrm{pop}}\left(t\right)\\
c_{\mathrm{EI}}^{\mathrm{pop}}\left(t\right) & c_{\mathrm{II}}^{\mathrm{pop}}\left(t\right) & c_{\mathrm{IX}}^{\mathrm{pop}}\left(t\right)\\
c_{\mathrm{EX}}^{\mathrm{pop}}\left(t\right) & c_{\mathrm{IX}}^{\mathrm{pop}}\left(t\right) & 0
\end{array}\right).
\]
In this way, the solutions $\delta\boldsymbol{m}(t)$ of \prettyref{eq:sigma_region}
are composed of all points that are one standard deviation away from
the expected activity. The covariances enter the total population
averaged variability, given by
\begin{eqnarray*}
c_{\alpha\beta}^{\mathrm{pop}}\left(t\right) & := & \left\langle \delta m_{\alpha}\left(t\right)\delta m_{\beta}\left(t\right)\right\rangle =\left\langle \frac{1}{N_{\alpha}}\sum_{i\in\alpha}\delta n_{i}\left(t\right)\frac{1}{N_{\beta}}\sum_{i\in\beta}\delta n_{i}\left(t\right)\right\rangle \\
 & = & \frac{\delta_{\alpha\beta}}{N_{\alpha}^{2}}\sum_{i\in\alpha}\left\langle \delta n_{i}^{2}\left(t\right)\right\rangle +\frac{1}{N_{\alpha}N_{\beta}}\sum_{i\in\alpha,\ j\in\beta,i\neq j}\left\langle \delta n_{i}\left(t\right)\delta n_{j}\left(t\right)\right\rangle \\
 & \approx & \delta_{\alpha\beta}\frac{a_{\alpha}\left(t\right)}{N_{\alpha}}+c_{\alpha\beta}\left(t\right)
\end{eqnarray*}
with the definitions \prettyref{eq:Def_a} and \prettyref{eq:Def_c_pairwise_pop}.}

\selectlanguage{american}%
The two points on the border of the dark gray error-ellipses of the
full covariances with the largest distance to the tangent of the limit
cycle at $\left(\left\langle \mI\right\rangle ,\left\langle \mE\right\rangle \right)$
are marked by a star, which, taken together, form the border of a
tube-shaped $\sigma$-area. This tube indicates the region in which
we most likely expect to find the system. To visualize the contributions
of auto- and pairwise covariances, we plot in light gray the error
ellipses based solely on the variances ($c^{\mathrm{pop}}\left(t\right)$
is diagonal in this case). The dark error ellipses are bigger than
the light ones, indicating that the covariances are positive and their
axes are tilted; the $c_{EI}=c_{IE}$-component is nonzero. Furthermore,
the error ellipses significantly change their size in time, indicative
of the modulation of the fluctuations with time. The variances grow
monotically with the respective mean activities, explaining that the
light gray ellipses are largest (smallest) where the mean activities
are largest (smallest). One can read off the phase shift of $c_{EE}$
to $m_{E}$ to be roughly $\frac{\pi}{2}$: the deviation of the dark
gray error ellipses from the light gray ones is largest at the points
where $m_{E}\left(t\right)\approx\overline{m}_{E}$ and $\delta m_{I}\left(t\right)$
is minimal. 

\selectlanguage{english}%

\selectlanguage{american}%

\section*{Methods\label{sec:Methods}}

\subsection*{Glauber dynamics in mean-field theory\label{subsec:The-mean-activity}}

We have left out so far several steps in the derivation of the results
that were not necessary for the presentation of the main ideas. In
this section, we will therefore give a self-contained derivation of
our results also necessitating paraphrases of some results known from
earlier works. The starting point is the master equation for the probability
density of the possible network states emerging from the Glauber dynamics
\citep{Glauber63_294} described in \nameref{subsec:Results_formulas}
(see for the following also \citep{Helias14,Buice09_377})

\begin{equation}
\frac{\partial p}{\partial t}(\boldsymbol{n},t)=\underbrace{\frac{1}{\tau}}_{\text{update rate}}\sum_{i=1}^{N}\underbrace{(2n_{i}-1)}_{\in\{-1,1\},\text{direction of flux}}\quad\underbrace{\phi_{i}(\boldsymbol{n}\backslash n_{i},t)}_{\text{net flux due to neuron \ensuremath{i}}}\quad\forall\quad\boldsymbol{n}\in\{0,1\}^{N},\label{eq:Master-equation}
\end{equation}
where
\begin{eqnarray*}
\phi_{i}(\boldsymbol{n}\backslash n_{i},t) & = & \underbrace{p(\boldsymbol{n}_{i-},t)\,F_{i}(\boldsymbol{n}_{i-})}_{\text{neuron }i\text{ transition up}}-\underbrace{p(\boldsymbol{n}_{i+},t)\,(1-F_{i}(\boldsymbol{n}_{i+}))}_{\text{neuron }i\text{ transition down}}\\
 & = & -p(\boldsymbol{n}_{i+})+p(\boldsymbol{n}_{i-},t)\,F_{i}(\boldsymbol{n}_{i-})+p(\boldsymbol{n}_{i+},t)\,F_{i}(\boldsymbol{n}_{i+}).
\end{eqnarray*}
The activation function $F_{i}\left(\boldsymbol{n}\right)$ is given
by \prettyref{eq:Def_gain_function}.

Using the master equation (for details cf. appendix, \prettyref{subsec:Derivation_of_moment_ODEs}),
one can derive a differential equation for the mean activity of the
neuron $i$, $\left\langle n_{i}\right\rangle \left(t\right)=\sum_{\boldsymbol{n}}p\left(\boldsymbol{n},t\right)n_{i}$
and the raw covariance of the neurons $i$ and $j$, $\left\langle n_{i}\left(t\right)n_{j}\left(t\right)\right\rangle =\sum_{\boldsymbol{n}}p\left(\boldsymbol{n},t\right)n_{i}n_{j}$
\citep{Renart10_587,Helias14,Ginzburg94,Glauber63_294,Buice09_377}.
This yields
\begin{align}
\tau\frac{d}{dt}\left\langle n_{k}\right\rangle \left(t\right) & =-\left\langle n_{k}\right\rangle \left(t\right)+\left\langle F_{k}\left(t\right)\right\rangle \label{eq:ODE_first_two_moments}\\
\frac{d}{dt}\left\langle n_{k}\left(t\right)n_{l}\left(t\right)\right\rangle  & =\left\{ -\left\langle n_{k}\left(t\right)n_{l}\left(t\right)\right\rangle +\left\langle n_{l}\left(t\right)F_{k}\left(t\right)\right\rangle \right\} +\left\{ k\leftrightarrow l\right\} .\nonumber 
\end{align}

As mentioned in \nameref{subsec:Results_formulas}, we assume that
the input $h_{i}$ coming from the local and the external population
is normally distributed, say with mean $\mu_{i}$ and standard deviation
$\sigma_{i}$ given by
\begin{align}
\mu_{i}(t) & :=\left\langle h_{i}\right\rangle =\left(J\left\langle \boldsymbol{n}\right\rangle \right)_{i}+\hext\sin(\omega t)\nonumber \\
\sigma_{i}^{2}(t) & :=\left\langle h_{i}^{2}\right\rangle -\left\langle h_{i}\right\rangle ^{2}=\sum_{k,k'=1}^{N}J_{i,k}J_{i,k'}\left(\left\langle n_{k}n_{k'}\right\rangle -\left\langle n_{k}\right\rangle \left\langle n_{k'}\right\rangle \right)+\left(\sigma_{i}^{\text{noise}}\right)^{2}\label{eq:Def_mu_sigma}\\
 & =\left(J^{T}cJ\right)_{ii}+J\circledast J\left\langle \boldsymbol{n}\right\rangle \circledast\left(1-\left\langle \boldsymbol{n}\right\rangle \right)+\left(\sigma_{i}^{\text{noise}}\right)^{2},\nonumber 
\end{align}
where the average $\langle\rangle$ is taken over realizations of
the stochastic dynamics and we used the element-wise (Hadamard) product
(see main text).

\begin{figure}[h]
\centering{}\includegraphics[scale=0.75]{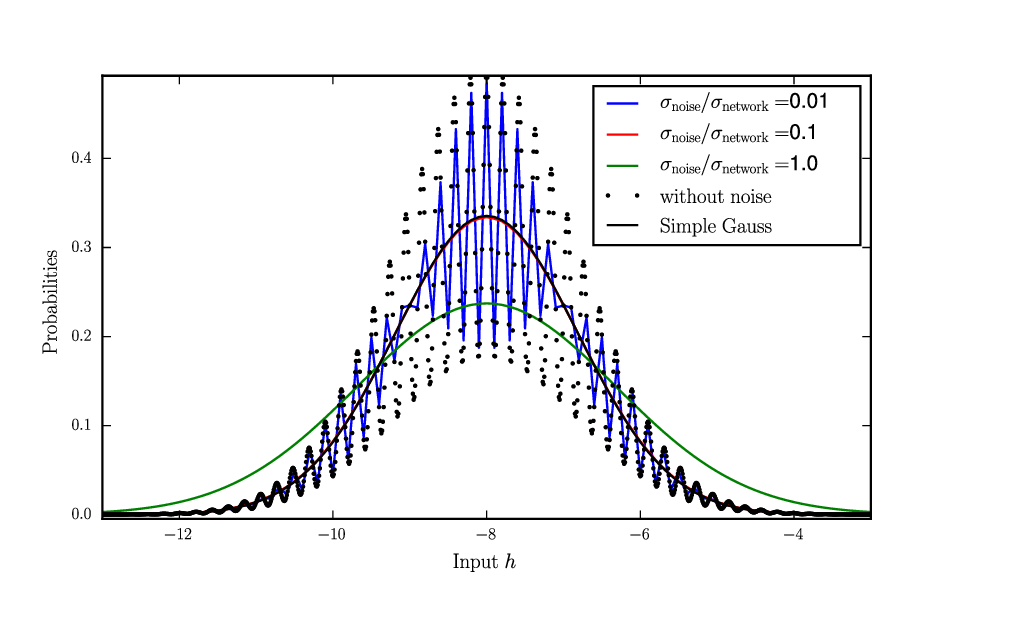}\caption{\foreignlanguage{american}{\textbf{Distribution of inputs from binary neurons for different noise
levels:} Probability distribution of synaptic input $h_{i}=\sum_{j}J_{ij}n_{j}+\xi_{i}$
of a neuron in a network of independently active cells $n_{j}$ with
$\langle n_{E}\rangle=\langle n_{I}\rangle=0.2$ and synaptic weights
$j_{I}=-0.21$, $j_{E}=0.01$. $\left|\frac{j_{E}}{j_{I}}\right|$
was deliberately chosen to be large because only then the convolution
of a binomial distribution ``squeezed'' to the step size $j_{E}$
with the binomial distribution squeezed to the step size $\left|j_{I}\right|$
results in a probability distribution with many local maxima leading
to the impression of an oscillation. The noiseless case $\xi_{i}=0$
is shown as black dots. The solid black curve indicates the Gaussian
approximation (cf. e.g. \prettyref{eq:Def_mu_sigma}, here without
perturbation) of this distribution from the main text. This distribution
appears in the expectation values of the activation function $F$
(cf. e.g. \prettyref{eq:Def_gain_function}): It is a Gaussian distribution
with the mean $\mu=K_{E}j_{E}m_{E}+K_{I}j_{I}m_{I}$ and the variance
$\sigma_{\text{network}}^{2}=K_{E}j_{E}^{2}m_{E}\left(1-m_{E}\right)+K_{I}j_{I}^{2}m_{I}\left(1-m_{I}\right)$
of the original binomial distributions $\mathrm{Binom(m_{E},K_{E})}$,
$\mathrm{Binom(m_{I},K_{I})}$. The other curves indicate convolutions
with the Gaussian noise $\xi\sim\mathcal{N}(0,\sigma_{\mathrm{noise}})$
of different magnitudes $\sigma_{\mathrm{noise}}$, given in units
of the noise level $\sigma_{\mathrm{network}}$ intrinsically produced
by the network.\label{fig:Noisy_input}}}
\end{figure}

The additional noise introduced in \prettyref{eq:Def_gain_function}
effectively leads to a smoothing of the neurons' activation threshold
and broadens the width of the input distribution. It can be interpreted
as additional variability coming from other brain areas. Furthermore,
it is computationally convenient, because the theory assumes the input
to be a (continuous) Gaussian distribution, while in the simulation,
the input $\sum_{l=k}^{N}J_{ik}n_{k}$, being a sum of discrete binary
variables, can only assume discrete values. The smoothing by the additive
noise therefore improves the agreement of the continuous theory with
the discrete simulation. Already weak external noise compared to the
intrinsic noise is sufficient to obtain a quite smooth probability
distribution of the input (\prettyref{fig:Noisy_input}).

The description in terms of a coupled set of moment equations instead
of the ODE for the full probability distribution here serves to reduce
the dimensionality: It is sufficient to describe the time evolution
of the moments on the population level, rather than on the level of
individual units. To this end we need to assume that the synaptic
weights $J_{ij}$ only depend on the population $\alpha,\ \beta\in\left\{ \mathrm{exc.},\ \mathrm{inh.},\ \mathrm{ext.}\right\} $
that $i$ and $j$ belong to, respectively, and thus (re)name them
$J_{\alpha\beta}$ (homogeneity). Furthermore, we assume that not
all neurons are connected to each other, but that $K_{\alpha\beta}$
is the number of incoming connections a neuron in population $\alpha$
receives from a neuron in population $\beta$ (fixed in-degree). The
incoming connections to each neuron are chosen randomly, uniformly
distributed over all possible sending neurons. This leads to expressions
for the population averaged input $h_{\alpha}$, mean activity $m_{\alpha}$
and covariance $c_{\alpha\beta}$, formally nearly identical to those
on the single cell level and analogous to those in \citep[sec. Mean-field solution]{Helias14}.

\selectlanguage{english}%

\selectlanguage{american}%

\subsubsection*{Mean activity: Stationary part and response to perturbation in linear
order}

We are now able to calculate the quantity $\left\langle F_{\alpha}\left(\boldsymbol{n}\left(t\right),t\right)\right\rangle =\left\langle H\left(h_{\alpha}\left(t\right)-\theta\right)\right\rangle $
(recall that $h_{\alpha}\left(t\right)$ is a Gaussian random variable
with mean $\mu_{\alpha}\left(t\right)$ and standard deviation $\sigma_{\alpha}\left(t\right)$),
the nonlinearity of the ODEs (\ref{eq:ODE_first_two_moments}) on
the population level. Multiplying $H\left(h_{\alpha}\left(t\right)-\theta_{\alpha}\right)$
by the Gaussian probability density for $h_{\alpha}\left(t\right)$,
we get, after substitution of the integration variable,
\begin{align}
 & \left\langle F_{\alpha}\left(\boldsymbol{n}\left(t\right),t\right)\right\rangle =\left\langle H\left(h_{\alpha}\left(t\right)-\theta_{\alpha}\right)\right\rangle \nonumber \\
= & \frac{1}{\sqrt{\pi}}\int_{\frac{\theta-\mu_{\alpha}\left(t\right)}{\sqrt{2}\sigma_{\alpha}\left(t\right)}}^{\infty}e^{-x^{2}}\,dx=\frac{1}{2}\text{erfc}\left(\frac{\theta_{\alpha}-\mu_{\alpha}\left(t\right)}{\sqrt{2}\sigma_{\alpha}\left(t\right)}\right)\nonumber \\
=: & \varphi(\mu_{\alpha}(\boldsymbol{m}\left(t\right),\hext\sin\left(\omega t\right))),\sigma{}_{\alpha}(\boldsymbol{m}\left(t\right),c\left(t\right))),\label{eq:Def_phi_pop}
\end{align}
where we defined the average input $\mu_{\alpha}$ and the width of
the input distribution $\sigma_{\alpha}$ 

\begin{eqnarray}
\mu_{\alpha}\left(t\right) & := & \left[\left(K\circledast J\right)\,\boldsymbol{m}\left(t\right)\right]_{\alpha}+\hext\text{sin}(\omega t)\label{eq:Def_mu_pop}\\
\sigma_{\alpha}^{2}\left(t\right) & := & \left[\left(K\circledast J\right)^{T}\,c\left(t\right)\,\left(K\circledast J\right)\right]_{\alpha\alpha}\label{eq:Def_sigma_pop}\\
 &  & +\left[K\circledast J\circledast J\,\boldsymbol{m}\left(t\right)\circledast\left(1-\boldsymbol{m}\left(t\right)\right)\right]_{\alpha}+\sigma_{\alpha,\text{\text{noise}}}^{2}.\nonumber 
\end{eqnarray}
Recall that we defined $\overline{x}$ to be the quantity $x$ in
the stationary case (without external input). For the linear approximation
around $\mu_{\alpha}=\overline{\mu}_{\alpha}$, $\sigma_{\alpha}=\overline{\sigma}_{\alpha}$
and $\hext=0$, we have to take into account all dependencies via
inner derivatives. We set $\delta\mu_{\alpha}=\mu_{\alpha}-\overline{\mu}_{\alpha}$
and $\delta\sigma_{\alpha}=\sigma_{\alpha}-\overline{\sigma}_{\alpha}$.
Note that $\delta\mu$ includes the variation of $\mu$ both because
of fluctuations in the network and because of the external drive.
The Taylor expansion up to linear order is
\[
\varphi(\mu{}_{\alpha}(\boldsymbol{m},c,\hext),\sigma_{\alpha}(\boldsymbol{m},c))\approx\varphi(\overline{\mu}_{\alpha},\overline{\sigma}_{\alpha})+\underbrace{S\left(\overline{\mu}_{\alpha},\overline{\sigma}_{\alpha}\right)}_{=:\overline{S}_{\alpha}}\left[\delta\mu_{\alpha}+\frac{\theta-\overline{\mu}_{\alpha}}{\overline{\sigma}_{\alpha}}\delta\sigma_{\alpha}\right],
\]
where we introduced the susceptibility on the population level
\begin{equation}
S\left(\mu_{\alpha}\left(t\right),\sigma_{\alpha}\left(t\right)\right):=\frac{d}{d\mu_{\alpha}\left(t\right)}\varphi\left(\mu_{\alpha}(t),\sigma{}_{\alpha}(t)\right)=\frac{1}{\sqrt{2\pi}\sigma_{\alpha}\left(t\right)}\,e^{-\frac{\left(\mu_{\alpha}\left(t\right)-\theta_{\alpha}\right)^{2}}{2\sigma_{\alpha}^{2}\left(t\right)}}.\label{eq:Def_susceptibility_pop}
\end{equation}
Now, we express $\delta\sigma_{\alpha}$ and $\delta\mu_{\alpha}$
via $\boldsymbol{\delta m}:=\boldsymbol{m}-\overline{\boldsymbol{m}}$
and $\delta c:=c-\overline{c}$ (cf. \citep[eq. (29)]{Helias14} for
the time-independent case): 
\begin{eqnarray}
\delta\mu_{\alpha}\left(t\right) & = & \sum_{\beta}K_{\alpha\beta}J_{\alpha\beta}\delta m_{\beta}\left(t\right)+\hext\sin\left(\omega t\right)\label{eq:Def_del_mu}\\
\delta\sigma_{\alpha}\left(t\right) & = & \frac{1}{2\sigma_{\alpha}}\left(\sum_{\beta}K_{\alpha\beta}J_{\alpha\beta}^{2}\left(1-2m_{\beta}\right)\delta m_{\beta}\left(t\right)+\sum_{\beta,\gamma}K_{\alpha\beta}K_{\alpha\gamma}J_{\alpha\beta}J_{\alpha\gamma}\delta c_{\beta\gamma}\left(t\right)\right).\label{eq:Def_del_sig}
\end{eqnarray}
Note that in $\delta\mu_{\alpha}$ (but not $\delta\sigma_{\alpha}$),
the perturbation occurs again explicitly. In \prettyref{eq:Proof_del_c_order_del_m_over_N},
we demonstrate that $\delta c$ scales like $\frac{\delta m}{N}$,
like in the stationary case. Furthermore, we certainly have $\frac{\left|K\right|}{N}=\mathcal{O}\left(1\right)$
and $\sigma=\mathcal{O}\left(\sqrt{\left|K\right|}\right)$, thus
\[
\left|\delta\boldsymbol{\mu}\left(t\right)\right|=\mathcal{O}\left(\left|K\right|\left|\delta\boldsymbol{m}\left(t\right)\right|\right)=\mathcal{O}\left(\hext\right),\text{ but }\left|\delta\boldsymbol{\sigma}\left(t\right)\right|=\mathcal{O}\left(\sqrt{\left|K\right|}\left|\delta m\left(t\right)\right|\right)=\mathcal{O}\left(\frac{\hext}{\sqrt{\left|K\right|}}\right).
\]
We therefore neglect $\delta\boldsymbol{\sigma}$ in our calculations
for $\delta\boldsymbol{m}$ from \prettyref{eq:ODE_mean_activity}
on. This yields for the linearization of the ODE \prettyref{eq:ODE_mean_population_averaged}
\begin{eqnarray}
\tau\frac{\partial}{\partial t}\delta m_{\alpha}\left(t\right)+\delta m_{\alpha}\left(t\right) & = & \overline{S}_{\alpha}\left[\delta\mu_{\alpha}\left(t\right)+\right.\underbrace{\frac{\theta_{\alpha}-\overline{\mu}_{\alpha}}{\overline{\sigma}_{\alpha}}}_{=\text{erfc}^{-1}\left(\bar{m}\right)}\left.\delta\sigma_{\alpha}\left(t\right)\right]\label{eq:ODE_mean_activity_incl_del_sigma}\\
\tau\frac{\partial}{\partial t}\delta m_{\alpha}\left(t\right)+\delta m_{\alpha}\left(t\right) & = & \sum_{\beta}W_{\alpha\beta}\delta m_{\beta}\left(t\right)+\overline{S}_{\alpha}\hext\sin\left(\omega t\right)+\mathcal{O}\left(\hext^{2},\frac{1}{\sqrt{K}}\right),\label{eq:ODE_mean_activity}
\end{eqnarray}
where we used the relation $\frac{\Theta_{\alpha}-\overline{\mu}_{\alpha}}{\overline{\sigma}_{\alpha}}=\sqrt{2}\mathrm{erfc}^{-1}\left(2\overline{m}_{\alpha}\right)$,
derived from \prettyref{eq:m_self_consistent} in connection with
\prettyref{eq:Def_phi_pop}, which implies that this expression does
not depend on $K$, but solely on $\overline{m}_{\alpha}$ and we
defined
\[
W_{\alpha\beta}:=\overline{S}_{\alpha}K_{\alpha\beta}J_{\alpha\beta}.
\]
The only change compared to the setup in \citep{Helias14} is again
the occurrence of a periodic term, here $S_{\alpha}\hext\sin\left(\omega t\right)$. 

We solve \prettyref{eq:ODE_mean_activity} by transforming it into
the eigenbasis of the matrix $W_{\alpha\beta}$ 
\begin{equation}
U^{-1}WU=\mathrm{\mathrm{diag}}\left(\lambda_{1},..,\lambda_{\tilde{N}}\right):=\Lambda.\label{eq:Diagonalise_W}
\end{equation}
We multiply \prettyref{eq:ODE_mean_activity} by $U^{-1}$, define
$\delta m^{\alpha}:=\left(U^{-1}\right)^{\alpha\beta}\delta m_{\beta}$
and get
\begin{equation}
\tau\frac{d}{dt}\delta m^{\alpha}=-\delta m^{\alpha}+\Lambda_{\beta}^{\alpha}\delta m^{\beta}+\left(U^{-1}\right)^{\alpha\beta}\overline{S}_{\beta}\hext\text{sin}\left(\omega t\right).\label{eq:epsilon_tilde}
\end{equation}
Note that the input is projected onto the respective eigenmodes. \prettyref{eq:epsilon_tilde}
can be solved including the transient phase by the method of variation
of constants. 

But as we are only interested in the cyclostationary part of the solution,
we can neglect the solution of the homogeneous part and solely compute
the particular solution. Observe that $\frac{d}{dt}\text{Im}(\delta m^{\alpha}(t))=\text{Im}\left(\frac{d}{dt}\delta m^{\alpha}(t)\right)$
for a differentiable function $\delta m^{\alpha}$ because $t\in\mathbb{R}$.
 We insert the ansatz $\delta m^{\alpha}=M_{1}^{\alpha}e^{i\omega t}$
and solve for $M_{1}^{\alpha}$, which gives \prettyref{eq:mean_activity_Fourier}
of the main text. For further calculations, keep in mind that $M_{\alpha}^{1}$
and therefore $\delta m_{\alpha}$ are of order $\mathcal{O}\left(\hext\frac{S}{SKJ}\right)=\mathcal{O}\left(\hext\frac{1}{KJ}\right)$.
In the appendix, \prettyref{subsec:Phase_extraction}, we describe
how to extract the right phase of the real solution from the complex
ansatz.

\subsubsection*{Covariances: Stationary part and response to a perturbation in linear
order\label{subsec:Methods_correlations}}

Using \prettyref{eq:ODE_first_two_moments} in the population-averaged
version, we calculate the derivative of the zero time-lag covariance
\[
c_{\alpha\beta}\left(t\right):=\frac{1}{N_{\alpha}N_{\beta}}\sum_{i\in\alpha,j\in\beta,i\neq j}\left\langle n_{i}\left(t\right)n_{j}\left(t\right)\right\rangle -\left\langle n_{i}\left(t\right)\right\rangle \left\langle n_{j}\left(t\right)\right\rangle 
\]
getting
\[
\tau\frac{dc_{\alpha\beta}\left(t\right)}{dt}=-2c_{\alpha\beta}\left(t\right)+\frac{1}{N_{\alpha}N_{\beta}}\sum_{i\in\alpha,j\in\beta,i\neq j}\left\langle F_{j}\left(\boldsymbol{n}\left(t\right)\right)\delta n_{i}\left(t\right)\right\rangle +\left\langle F_{i}\left(\boldsymbol{n}\left(t\right)\right)\delta n_{j}\left(t\right)\right\rangle .
\]
Neglecting cumulants of order higher than two, we can expand the expectation
value $\left\langle F_{i}\left(\boldsymbol{n}\left(t\right)\right)\delta n_{j}\left(t\right)\right\rangle $
(cf. \citep[section "Linearized equation for correlations and susceptibility"]{Helias14,Buice10_377})
and get

\begin{equation}
\left\langle F_{i}\left(\boldsymbol{n}\left(t\right)\right)\delta n_{j}\left(t\right)\right\rangle \approx S\left(\mu_{i}\left(t\right),\sigma_{i}\left(t\right)\right)\,\sum_{k\neq j}J_{ik}c_{kj}\left(t\right)+S\left(\mu_{i}\left(t\right),\sigma_{i}\left(t\right)\right)\,J_{ij}a_{j}\left(t\right).\label{eq:ODE_correlations_general_indices}
\end{equation}
After carrying out the population averaging, we get the ordinary differential
equation
\begin{align}
 & \tau\frac{dc_{\alpha\beta}\left(t\right)}{dt}\nonumber \\
= & \left\{ -c_{\alpha\beta}\left(t\right)+\sum_{\gamma}S\left(\mu_{\alpha}\left(t\right),\sigma_{\alpha}\left(t\right)\right)\,K_{\alpha\gamma}J_{\alpha\gamma}\left(c_{\gamma\beta}\left(t\right)+\delta_{\gamma\beta}\frac{a_{\beta}\left(t\right)}{N_{\beta}}\right)\right\} +\left\{ \alpha\leftrightarrow\beta\right\} .\label{eq:ODE_full_c}
\end{align}

\selectlanguage{english}%
\foreignlanguage{american}{Therefore, the stationary part $\overline{c}$
of the covariances fulfills the relation (cf. \citep{Helias14,Buice10_377})
\begin{equation}
2\overline{c}_{\alpha\beta}=\sum_{\gamma}S\left(\overline{\mu}_{\alpha},\overline{\sigma}_{\alpha}\right)\left(K\circ J\right)_{\alpha\gamma}\left(\overline{c}_{\gamma\beta}+\delta_{\gamma\beta}\frac{\overline{a}_{\beta}}{N_{\beta}}\right)+\alpha\leftrightarrow\beta.\label{eq:Stationary_correlations}
\end{equation}
As for the mean activities, we want to make a little step (of order
$\hext$, to be precise) away from the stationary state determining
the deviation $\delta c\left(t\right):=c\left(t\right)-\overline{c}$.
For that, we have to calculate the Taylor expansion of $S\left(\mu_{\alpha}\left(t\right),\sigma_{\alpha}\left(t\right)\right)$
in $\delta\boldsymbol{m}$, i.e.
\begin{align*}
 & S\left(\mu_{\alpha}\left(t\right),\sigma_{\alpha}\left(t\right)\right)\\
:= & \frac{1}{\sqrt{2\pi}}\frac{1}{\sigma_{\alpha}\left(t\right)}\text{exp}\left(-\frac{\left(\mu_{\alpha}\left(t\right)-\theta_{\alpha}\right)^{2}}{2\left(\sigma_{\alpha}\left(t\right)\right)^{2}}\right)\\
\approx & S\left(\overline{\mu}_{\alpha},\overline{\sigma}_{\alpha}\right)+\left(\frac{\partial S}{\partial\mu_{\alpha}\left(t\right)}\delta\mu_{\alpha}+\frac{\partial S}{\partial\sigma_{\alpha}\left(t\right)}\delta\sigma_{\alpha}\right)\bigg\vert_{\mu_{\alpha}=\overline{\mu}_{\alpha},\sigma_{\alpha}=\overline{\sigma}_{\alpha}},
\end{align*}
 where $\delta\mu_{\alpha}$ and $\delta\sigma_{\alpha}$ are given
by \prettyref{eq:Def_del_mu} and 
\begin{eqnarray*}
\frac{\partial S}{\partial\mu_{\alpha}\left(t\right)}\left(\overline{\mu}_{\alpha},\overline{\sigma}_{\alpha}\right) & = & \frac{\theta_{\alpha}-\overline{\mu}_{\alpha}}{\overline{\sigma}_{\alpha}^{2}}S\left(\overline{\mu}_{\alpha},\overline{\sigma}_{\alpha}\right)\\
\frac{\partial S}{\partial\sigma_{\alpha}\left(t\right)}\left(\overline{\mu}_{\alpha},\overline{\sigma}_{\alpha}\right) & = & -\frac{1}{\overline{\sigma}_{\alpha}}\left(1-\left(\frac{\theta_{\alpha}-\overline{\mu}_{\alpha}}{\overline{\sigma}_{\alpha}}\right)^{2}\right)S\left(\overline{\mu}_{\alpha},\overline{\sigma}_{\alpha}\right)\\
 & = & \frac{\theta_{\alpha}-\overline{\mu}_{\alpha}}{\overline{\sigma}_{\alpha}^{2}}\left(\underbrace{\frac{\theta_{\alpha}-\overline{\mu}_{\alpha}}{\overline{\sigma}_{\alpha}}}_{=\mathcal{O}\left(1\right)}-\underbrace{\frac{\overline{\sigma}_{\alpha}}{\theta_{\alpha}-\overline{\mu}_{\alpha}}}_{=\mathcal{O}\left(1\right)}\right)S\left(\overline{\mu}_{\alpha},\overline{\sigma}_{\alpha}\right)
\end{eqnarray*}
Here again, the relation $\frac{\Theta_{\alpha}-\overline{\mu}_{\alpha}}{\overline{\sigma}_{\alpha}}=\sqrt{2}\mathrm{erfc}^{-1}\left(2\overline{m}_{\alpha}\right)$
was used to estimate the dependence on $K$. We insert the linearization
of $S$ and the expressions for $\delta\mu$ and $\delta\sigma$,
\prettyref{eq:Def_del_mu}, into the ODE for $c_{\alpha\beta}\left(t\right)=\overline{c}_{\alpha\beta}+\delta c_{\alpha\beta}\left(t\right)$
to get, after neglecting the contributions of order $\mathcal{O}\left(\hext^{2}\right)$
and sorting the rest into terms proportional to $\delta c$, $\hext$
and $\delta m$ respectively:
\begin{align}
 & \tau\frac{d}{dt}\delta c_{\alpha\beta}\left(t\right)+\left\{ \sum_{\gamma}\left(\delta_{\alpha\gamma}-S\left(\overline{\mu}_{\alpha},\overline{\sigma}_{\alpha}\right)K_{\alpha\gamma}J_{\alpha\gamma}\right)\delta c_{\gamma\beta}\left(t\right)\right\} +\left\{ \alpha\leftrightarrow\beta\right\} \nonumber \\
= & \left\{ \frac{\partial S}{\partial\mu_{\alpha}\left(t\right)}\sum_{\gamma}K_{\alpha\gamma}J_{\alpha\gamma}\left(\overline{c}{}_{\gamma\beta}+\frac{\overline{a}_{\beta}}{N_{\beta}}\delta_{\gamma\beta}\right)\right.\Big(\hext\sin\left(\omega t\right)+\underbrace{\sum_{\delta}K_{\alpha\delta}J_{\alpha\delta}\delta m_{\delta}\left(t\right)}_{=\mathcal{O}\left(\hext\right)}\Big)\label{eq:ODE_c_first_version}\\
+ & \frac{\partial S}{\partial\sigma_{\alpha}\left(t\right)}\sum_{\gamma}K_{\alpha\gamma}J_{\alpha\gamma}\left(\overline{c}_{\gamma\beta}+\frac{\overline{a}_{\beta}}{N_{\beta}}\delta_{\gamma\beta}\right)\underbrace{\delta\sigma_{\alpha}\left(t\right)}_{=\mathcal{O}\left(\frac{\hext}{\sqrt{K}}\right)}\nonumber \\
+ & \left.S\left(\overline{\mu}_{\alpha},\overline{\sigma}_{\alpha}\right)K_{\alpha\beta}J_{\alpha\beta}\frac{\left(1-2\overline{m}{}_{\beta}\right)}{N_{\beta}}\delta m_{\beta}\left(t\right)\right\} +\left\{ \alpha\leftrightarrow\beta\right\} \nonumber 
\end{align}
Before finally solving for $\delta c\left(t\right)$, we want to justify
the assumption $\delta c=\mathcal{O}\left(\frac{\delta m}{N}\right)$,
which we needed already in the beginning to determine $\delta m\left(t\right)$,
by a short calculation. We insert \prettyref{eq:Def_del_sig} into
\prettyref{eq:ODE_c_first_version} and switch to matrix notation
for brevity, which yields
\begin{align}
 & \tau\frac{d}{dt}\delta c\left(t\right)+\left\{ \left(\mathbbm{1}-\overline{S}KJ\right)\delta c\left(t\right)\right\} +\left\{ ...\right\} ^{T}\nonumber \\
= & \Bigg\{\frac{\partial S}{\partial\mu}KJ\left(\overline{c}+\frac{\overline{a}}{N}\right)\left(\hext\sin\left(\omega t\right)+KJ\delta m\left(t\right)\right)+\underbrace{\frac{\partial S}{\partial\sigma}}_{=\frac{\partial S}{\partial\mu}\left(\frac{\Theta-\mu}{\sigma}-\frac{\sigma}{\Theta-\mu}\right)}KJ\left(\overline{c}+\frac{\overline{a}}{N}\right)\nonumber \\
 & \times\left(\frac{KJ^{2}\left(1-2\overline{m}\right)}{2\overline{\sigma}}\delta m\left(t\right)+\frac{KJ}{2\sigma}\delta c\left(t\right)\left(KJ\right)^{T}\right)+\overline{S}KJ\frac{1-2\overline{m}}{N}\delta m\left(t\right)\Bigg\}+\left\{ ...\right\} ^{T}.\label{eq:ODE_del_c_more_explicit}
\end{align}
We can rewrite the left hand side in order to recognize the parts,
which are identical to the right hand side of \prettyref{eq:ODE_mean_activity_incl_del_sigma},
i.e. the ODE for $\delta m\left(t\right)$ without the neglect of
$\delta\sigma$, which gives
\begin{align}
 & \Bigg\{\frac{\partial S}{\partial\mu}KJ\left(\overline{c}+\frac{\overline{a}}{N}\right)\label{eq:ODE_for_del_m_in_del_c}\\
 & \times\underbrace{\left(\left(\hext\sin\left(\omega t\right)+KJ\delta m\left(t\right)\right)+\frac{\Theta-\mu}{\sigma}\left(\frac{KJ^{2}\left(1-2\overline{m}\right)}{2\overline{\sigma}}\delta m\left(t\right)+\frac{KJ}{2\sigma}\delta c\left(t\right)\left(KJ\right)^{T}\right)\right)}_{=\left(\tau\frac{\partial}{\partial t}\delta m\left(t\right)+\delta m\left(t\right)\right)/S}\nonumber \\
 & -\underbrace{\frac{\partial S}{\partial\mu}}_{=\frac{1}{\overline{\sigma}}\frac{\Theta-\overline{\mu}}{\overline{\sigma}}\overline{S}}\frac{\sigma}{\Theta-\mu}KJ\left(\overline{c}+\frac{\overline{a}}{N}\right)\left(\frac{KJ^{2}\left(1-2\overline{m}\right)}{2\overline{\sigma}}\delta m\left(t\right)+\frac{KJ}{2\sigma}\delta c\left(t\right)\left(KJ\right)^{T}\right)\nonumber \\
 & +\overline{S}KJ\frac{1-2\overline{m}}{N}\delta m\left(t\right)\Bigg\}+\left\{ ...\right\} ^{T}.\nonumber 
\end{align}
Bringing the $\delta c$-terms on the left hand side finally yields
\begin{align}
 & \tau\frac{d}{dt}\delta c\left(t\right)+\big\{\left(\mathbbm{1}-\overline{S}KJ\right)\delta c\left(t\right)+\underbrace{\frac{1}{\overline{\sigma}}\overline{S}KJ\left(\overline{c}+\frac{\overline{a}}{N}\right)\frac{KJ}{2\sigma}\delta c\left(t\right)\left(KJ\right)^{T}}_{=\mathcal{O}\left(\overline{S}KJ\frac{K}{N}\frac{KJ^{2}}{\sigma^{2}}\delta c\left(t\right)\right)=\mathcal{O}\left(\overline{S}KJ\frac{K}{N}\delta c\left(t\right)\right)}\big\}+\left\{ ...\right\} ^{T}\nonumber \\
= & \Bigg\{\underbrace{\underbrace{\frac{\frac{\partial S}{\partial\mu}}{S}}_{=\frac{1}{\sigma}\frac{\Theta-\overline{\mu}}{\overline{\sigma}}}KJ\left(\overline{c}+\frac{\overline{a}}{N}\right)\left(\tau\frac{\partial}{\partial t}\delta m_{\alpha}\left(t\right)+\delta m_{\alpha}\left(t\right)\right)}_{=\mathcal{O}\left(\overline{S}KJ\frac{1}{N}\delta m\left(t\right)\right)}\label{eq:Proof_del_c_order_del_m_over_N}\\
 & -\underbrace{\frac{1}{\overline{\sigma}}\overline{S}KJ\left(\overline{c}+\frac{\overline{a}}{N}\right)\frac{KJ^{2}\left(1-2\overline{m}\right)}{2\overline{\sigma}}\delta m\left(t\right)}_{=\mathcal{O}\left(\overline{S}KJ\frac{1}{N}\delta m\left(t\right)\right)}+\overline{S}KJ\frac{1-2\overline{m}}{N}\delta m\left(t\right)\Bigg\}+\left\{ ...\right\} ^{T}.\nonumber 
\end{align}
We have therefore shown that - independent of the scaling of the synaptic
weights $J$ - the relation $\delta c=\mathcal{O}\left(\frac{\delta m}{N}\right)$
holds not only for the zero-mode, i.e. for the stationary case, but
also for the time-dependent part. Note that for our actual calculation
of $\delta m$, we have neglected its dependence on $\delta\sigma$,
as it is one order $\sqrt{K}$ smaller than the $\delta\mu$-contribution.
However, this is not true for $\delta c$ because of the cancellation
of the two contributions to $\delta\mu$. Inserting the rhs of the
ODE \prettyref{eq:ODE_mean_activity} actually used to determine $\delta m$
and shifting the $\delta c$-contribution of $\delta\sigma$ back
to the other side, we arrive at
\begin{align}
 & \tau\frac{d}{dt}\delta c\left(t\right)+\left\{ \left(\mathbbm{1}-\overline{S}KJ\right)\delta c\left(t\right)\right\} +\left\{ ...\right\} ^{T}\nonumber \\
= & \left\{ \frac{\frac{\partial S}{\partial\mu}}{S}KJ\left(\overline{c}+\frac{\overline{a}}{N}\right)\left(\tau\frac{\partial}{\partial t}\delta m_{\alpha}\left(t\right)+\delta m_{\alpha}\left(t\right)\right)\right.\nonumber \\
 & -\frac{1}{\overline{\sigma}}\left(1-\left(\frac{\mu-\Theta}{\sigma}\right)^{2}\right)\overline{S}KJ\left(\overline{c}+\frac{\overline{a}}{N}\right)\left(\frac{KJ^{2}\left(1-2\overline{m}\right)}{2\overline{\sigma}}\delta m\left(t\right)+\frac{KJ}{2\sigma}\delta c\left(t\right)\left(KJ\right)^{T}\right)\label{eq:ODE_delta_c_delta_sigma_on_one_side}\\
 & \left.+\overline{S}KJ\frac{1-2\overline{m}}{N}\delta m\left(t\right)\right\} +\left\{ ...\right\} ^{T}.\nonumber 
\end{align}
We want to compare the contribution from $\delta\mu$ in the second
line of \prettyref{eq:ODE_delta_c_delta_sigma_on_one_side} with the
contribution from $\delta\sigma$ in the third line. As pointed out
above, they scale in the same way with the system size $N$, given
that we do not rescale the driving frequency with $N$. Therefore,
its contribution stays equally important if we enlarge the network.
We neglect it anyway, which can be justified by comparing the decisive
part of the prefactors of the $\delta\sigma$ and the $\delta\mu$-parts
(the remaining parts are of the same order of magnitude):
\[
\overline{\sigma}\frac{\frac{\partial S}{\partial\mu}}{S}=\frac{\Theta-\overline{\mu}}{\overline{\sigma}}=\sqrt{2}\mathrm{erfc}^{-1}\left(2\overline{m}\right)\overset{\text{for input fluct. not too small}}{\gg}\frac{1}{\overline{\sigma}}\exp\left(-\mathrm{erfc}^{-1}\left(2\overline{m}\right)^{2}\right)=\overline{S}.
\]
This inequality is fulfilled for the three settings used in this work,
whereas the first term is one or two orders of magnitude larger than
the second. Especially, this inequality can always be fulfilled if
the externally generated noise level is high. Therefore, even if the
neglect of the $\delta\sigma$-contribution to $\delta c$ cannot
be justified by the standard mean-field argument that it decays faster
with the system size than other terms, it is applicable because the
input fluctuations are large enough - for all system sizes. This largely
simplifies the calculations because the ODE for $\delta c$ can be
solved by transforming into the eigensystem of $W$, which would not
be possible after including the more involved term emerging from $\delta\sigma$.
Taking into account the neglected term would require to reformulate
the problem as an equation for the vector $\left(\delta c_{\mathrm{EE}},\delta c_{\mathrm{EI}},..\right)$,
which would be much less intuitive. Furthermore, there is an indirect
argument for high frequencies that does rely on the system size: The
$\omega$-dependence of the absolute value of the maxima of $\delta m$
and $\delta c$ scales with the eigenvalues of $W$, which scale with
$\sqrt{K}$. Thus, changing the system size $N$ in first order just
stretches the $\omega$-axis. Therefore, the ``interesting'' frequencies
do scale with $N$, which leads to the dominance of the derivative
term in the second line of \prettyref{eq:ODE_delta_c_delta_sigma_on_one_side}
over the $\delta\sigma$-term. Note that the observation from \prettyref{eq:ODE_for_del_m_in_del_c}
that
\begin{align}
 & \underbrace{\left(KJ\left(1+\mathcal{O}\left(\frac{1}{\sqrt{K}}\right)\right)\delta m\right)^{\text{diag}}}_{=\mathcal{O}\left(\hext\right)}+\underbrace{\hext\sin\left(\omega t\right)}_{=\mathcal{O}\left(\hext\right)}\label{eq:}\\
= & \left(\overline{S}^{\text{diag}}\right)^{-1}\left(\tau\frac{\partial}{\partial t}\delta m+\delta m\right)+\mathcal{O}\left(\sqrt{\left|K\right|}\left|\delta\boldsymbol{m}\right|\right)=\mathcal{O}\left(\sqrt{\left|K\right|}\left|\delta\boldsymbol{m}\right|\right)=\mathcal{O}\left(\frac{\hext}{\sqrt{\left|K\right|}}\right)\label{eq:ODE_corr_magnitudes_estimate}
\end{align}
is a direct consequence of the recurrent drive being effectively inhibitory
(for other networks, the expansion around the stationary point would
not make sense): Any of the two terms in the susceptibility terms
are of order $\sqrt{K}$ bigger than their sum. Furthermore, we see
from \prettyref{eq:Proof_del_c_order_del_m_over_N} that the sum of
the susceptibility terms is of the same order of magnitude with respect
to its dependence on $K$ (or, equivalently, the connection probabilities
and the system size) as the term coming from the time modulation of
the variances (modulated-autocovariances-drive). }

\selectlanguage{american}%
We define\foreignlanguage{english}{ }
\begin{eqnarray}
T_{\alpha\beta} & := & K_{\alpha\beta}J_{\alpha\beta}\nonumber \\
V_{\alpha\beta} & := & \frac{\Theta-\overline{\mu}_{\alpha}}{\left(\overline{\sigma}_{\alpha}\right)^{2}}S\left(\overline{\mu}_{\alpha},\overline{\sigma}_{\alpha}\right)K_{\alpha\beta}J_{\alpha\beta},\label{eq:Def_T_V-1}
\end{eqnarray}
and
\begin{eqnarray}
N_{\alpha\beta}^{\mathrm{diag}} & = & \delta_{\alpha\beta}N_{\alpha}\nonumber \\
\overline{m}_{\alpha\beta}^{\mathrm{diag}} & = & \delta_{\alpha\beta}\overline{m}_{\alpha}\nonumber \\
\overline{a}_{\alpha\beta}^{\mathrm{diag}} & = & \delta_{\alpha\beta}\overline{a}_{\alpha}\nonumber \\
\delta m_{\alpha\beta}^{\mathrm{diag}}\left(t\right) & = & \delta_{\alpha\beta}\delta m_{\alpha}\left(t\right)\label{eq:Def_diag-1}\\
\left(T\delta m\left(t\right)\right)_{\alpha\beta}^{\mathrm{diag}} & = & \delta_{\alpha\beta}\sum_{\gamma}T_{\alpha\gamma}\delta m_{\gamma}\left(t\right),\nonumber 
\end{eqnarray}
we end up with the index-free version \prettyref{eq:ODE_compact_coloured}.
The first two inhomogeneities, the susceptibility terms introduced
in the main part (\nameref{sec:Results}) reflect the nonlinearity
of the gain-function.

With $U$ given in \prettyref{eq:Diagonalise_W}, we multiply \prettyref{eq:ODE_c_first_version}
from the left by $U^{-1}$ and from the right by $\left(U^{-1}\right)^{T}$
to get (cf. \citep{Buice10_377,Helias14}) 
\begin{eqnarray*}
 &  & \tau\frac{d}{dt}\underbrace{U^{-1}\delta c\left(t\right)\left(U^{-1}\right)^{T}}_{:=\widetilde{\delta c}\left(t\right)}\\
 & = & \left\{ \left(-\mathbbm{1}+\right.\right.\underbrace{U^{-1}WU}_{=\Lambda}\left.\right)\underbrace{U^{-1}\delta c\left(t\right)\left(U^{-1}\right)^{T}}_{:=\widetilde{\delta c}\left(t\right)}\\
 & + & U^{-1}\left(\left(T\delta m\left(t\right)\right)^{\text{diag}}+\hext\sin\left(\omega t\right)\right)V\left(\overline{c}+\frac{1}{N^{\text{diag}}}\overline{a}^{\text{diag}}\right)\left(U^{-1}\right)^{T}\\
 & + & U^{-1}W\left(\mathbbm{1}-2\overline{m}^{\text{diag}}\right)\frac{1}{N^{\text{diag}}}\left.\delta m\left(t\right)^{\text{diag}}\left(U^{-1}\right)^{T}\right\} \\
 & + & \left\{ ...\right\} ^{T}.
\end{eqnarray*}
We are only interested in the cyclostationary statistics, so we can
ignore again the transient state making the ansatz $\widetilde{\delta c_{\alpha\beta}^{\text{inhom}}}=\widetilde{C_{\alpha\beta}^{1}}e^{i\omega t}$.
Inserting this ansatz and transforming back into the original system,
we get
\begin{align}
\widetilde{C_{\alpha\beta}^{1}} & =\hext\frac{-i\tau\omega+2-\left(\lambda_{\alpha}+\lambda_{\beta}\right)}{\left(\tau\omega\right)^{2}+\left(2-\left(\lambda_{\alpha}+\lambda_{\beta}\right)\right)^{2}}\nonumber \\
 & \left[\sum_{\gamma,\delta,\theta,\phi,\eta}U_{\alpha\eta}^{-1}U_{\beta\delta}^{-1}T_{\eta\theta}U_{\theta\phi}\left(U^{-1}S\right)_{\phi}\frac{-i\tau\omega+1-\lambda_{\phi}}{\left(\tau\omega\right)^{2}+\left(1-\lambda_{\phi}\right)^{2}}V_{\eta,\gamma}\left(\overline{c}+\frac{1}{N^{\text{diag}}}\overline{a}^{\text{diag}}\right)_{\gamma\delta}\right.\label{eq:Correlation_Fourier}\\
 & +\sum_{\gamma,\delta,\epsilon}U_{\alpha\epsilon}^{-1}U_{\beta\delta}^{-1}V_{\epsilon\gamma}\left(\overline{c}+\frac{1}{N^{\text{diag}}}\overline{a}^{\text{diag}}\right)_{\gamma\delta}\\
 & +\left.\sum_{\theta,\phi,\gamma}U_{\alpha\gamma}^{-1}U_{\beta\theta}^{-1}W_{\gamma\theta}\left(1-2\overline{m}_{\theta}^{\text{diag}}\right)\frac{1}{N_{\theta}}U_{\theta\phi}\left(U^{-1}S\right)_{\phi}\frac{-i\tau\omega+1-\lambda_{\phi}}{\left(\tau\omega\right)^{2}+\left(1-\lambda_{\phi}\right)^{2}}\right]\nonumber 
\end{align}
Together with \prettyref{eq:mean_activity_Fourier}, this is the main
result of this section.

\section*{Discussion}

The present work offers an extension of the well-known binary neuronal
network model beyond the stationary case \citep{Renart10_587,Helias14,Ginzburg94,VanVreeswijk98_1321,Buice10_377}.
We here describe the influence of a sinusoidally modulated input on
the mean activities and the covariances to study the statistics of
recurrently generated network activity in an oscillatory regime, ubiquitously
observed in cortical activity \citep{Buzsaki04_1926}.

Comparing with the results of the simulation of the binary network
with NEST \citep{Gewaltig_07_11204,Nest280} and the numerical solution
of the full mean-field ODE, we are able to show that linear perturbation
theory is sufficient to explain the most important effects occurring
due to sinusoidal drive. This enables us to understand the mechanisms
by the help of analytical expressions and furthermore we can predict
the network response to any time-dependent perturbation with existing
Fourier representation by decomposing the perturbing input into its
Fourier components.

We find that the amplitude of the modulation of the mean activity
is of the order $\hext/\left(\left(1-\lambda_{\alpha}\right)^{2}+\left(\tau\omega\right)^{2}\right)^{\frac{1}{2}}$,
where $\lambda_{\alpha},\alpha\in\left\{ E,I\right\} $ are the eigenvalues
of the effective connectivity matrix $W$, i.e. the input is filtered
by a first order low-pass filter and the amplitude of the modulation
decays like $\propto\omega^{-1}$ for large frequencies. This finding
is in line with earlier work on the network susceptibility \citep[esp. section V]{Ginzburg94}.

The qualitatively new result here is the identification of two distinct
mechanisms by which covariances $\delta c$ are modulated in time.
First, covariances are driven by the direct modulation of the susceptibility
$S$ due to the time-dependent external input and by the recurrent
input from the local network. Second, time-modulated variances, analogous
to their role in the stationary setting \citep{Helias14}, drive the
pairwise covariances.

Our setup is the minimal network model, in which these effects can
be observed - minimal in the sense that we would lose these properties
if we further simplified the model: The presence of a nonlinearity
in the neuronal dynamics, here assumed to be a threshold-like activation
function, is required for the modulation of covariances by the time-dependent
change of the effective gain. In a linear rate model \citep{Tetzlaff12_e1002596,Grytskyy13_131}
this effect would be absent, because mean activities and covariances
then become independent.

The second mechanism relies on the binary nature of neuronal signal
transmission: the variance $a(t)$ of the binary neuronal signal is,
at each point in time, completely determined by its mean $m(t)$.
This very dependence provides the second mechanism by which the temporally
modulated mean activity causes time-dependent covariances, because
all fluctuations and therefore all covariances are driven by the variance
$a(t)$.

Rate models have successfully been used to explain the smallness of
pairwise covariances \citep{Renart10_587} by negative feedback \citep{Tetzlaff12_e1002596}.
A crucial difference is that their state is continuous, rather than
binary. As a consequence, the above-mentioned fluctuations present
due to the discrete nature of the neuronal signal transmission need
to be added artificially: The pairwise statistics of spiking or binary
networks are equivalent to the statistics of rate models with additive
white noise \citep{Grytskyy13_131}. To obtain qualitative or even
quantitative agreement of time-dependent covariances between spiking
or binary networks and rate models, the variance of this additive
noise needs to be chosen such that its variance is a function of the
mean activity and its time derivative. 

The direct modulation of the susceptibility $S$ due to the time-dependent
external input leads to a contribution to the covariances with first
order low-pass filter characteristics that dominates the modulated
covariances at large frequencies. For small - and probably biologically
realistic - frequencies (typically the LFP shows oscillations in the
$\beta$-range around $20\,\mathrm{Hz}$), however, the modulation
of the susceptibility by the local input from the network leads to
an equally important additional modulation of the susceptibility.
The intrinsic fluctuations of the network activity are moreover driven
by the time-dependent modulation of the variance, which is a function
of the mean activity as well. Because the mean activity follows the
external drive in a low-pass filtered manner, the latter two contributions
hence exhibit a second order low-pass-filter characteristics. These
contributions are therefore important at the small frequencies we
are interested in here.

The two terms modulating the susceptibility, by the direct input and
by the feedback of the mean activity through the network, have opposite
signs in balanced networks. In addition they have different frequency
dependencies. In networks in which the linearized connectivity has
only real eigenvalues, these two properties together lead to their
summed absolute value having a maximum. Whether or not the total modulation
of the covariance shows resonant behavior, however, depends also on
the third term that stems from the modulated variances. We find that
in purely inhibitory networks, the resonance peak is typically overshadowed
by the latter term. This is because inhibitory feedback leads to negative
average covariances \citep{Helias14}, which we show here reduce the
driving force for the two resonant contributions. In balanced E-I
networks, the driving force is not reduced, so the resonant contribution
can become dominant.

For the biologically motivated parameters used in the last setting
studied here, the effective coupling matrix $W$ has complex eigenvalues
which cause resonant mean activities. If the inhomogeneity was independent
of the driving frequency, $\delta c$ would have resonant modes with
frequency $f_{\mathrm{res}}$ and $2f_{\mathrm{res}}$. Due to the
mixing of the different modes and by the frequency dependence of the
inhomogeneity driving the modulation of covariances, these modes determine
only the ballpark for the location of the resonance in the covariance.
Especially the resonances are not sharp enough so that each of them
is visible in any combination of the modes. Different behavior is
expected near the critical point where $\Re\left(\lambda\right)\lesssim1$.

For predictions of experimental results, however, a more careful choice
of reasonable biological parameters would be necessary. In particular,
the external drive should be gauged such that the modulations of the
mean activities are in the experimentally observed range. Still, our
setup shows that the theory presented here works in the biologically
plausible parameter range.

The goal of extracting fundamental mechanisms of time-dependent covariances
guides the here presented choice of the level of detail of our model.
Earlier works \citep{Renart10_587,VanVreeswijk98_1321,Vreeswijk96}
showed that our setup without sinusoidal drive is sufficient to qualitatively
reproduce and explain phenomena observed in vivo, like high variability
of neuronal activity and small covariances. The latter point can be
explained in binary networks by the suppression of fluctuations by
inhibitory feedback, which is a general mechanism also applicable
to other neuron models \citep{Tetzlaff12_e1002596} and even finds
application outside neuroscience, for example in electrical engineering
\citep{Oppenheim96}. The high variability observed in binary networks
can be explained by the network being in the balanced state, that
robustly emerges in the presence of negative feedback \citep{Vreeswijk96,Amit-1997_373}.
In this state, the mean excitatory and inhibitory synaptic inputs
cancel so far that the summed input to a neuron fluctuates around
its threshold. This explanation holds also for other types of model
networks and also for biological neural networks \citep{Shadlen94}.
We have seen here that the operation in the balanced state, at low
frequencies, gives rise to a partial cancellation of the modulation
of covariances.

Our assumption of a network of homogeneously connected binary neurons
implements the general feature of neuronal networks that every neuron
receives input from a macroscopic number of other neurons, letting
the impact of a single synaptic afferent on the activation of a cell
be small and the summed input be distributed close to Gaussian: For
uncorrelated incoming activity, the ratio between the fluctuations
caused by a single input and the fluctuations of the total input is
$N^{-\frac{1}{2}}$, independent of how synapses scale with $N$.
However, the input to a neuron is actually not independent, but weakly
correlated, with covariances decaying at least as fast as $N^{-1}$
\citep{Renart10_587,Vreeswijk96}. Therefore this additional contribution
to the fluctuations also decays like $N^{-\frac{1}{2}}$. The Gaussian
approximation of the synaptic input relies crucially on these properties.
Dahmen et al. \citep{Dahmen16_031024} investigated third order cumulants,
the next order of non-Gaussian corrections to this approximation.
They found that the approximation has a small error even down to small
networks of about $500$ neurons and $50$ synaptic inputs per neuron.
These estimates hold as long as all synaptic weights are of equal
size. For distributed synaptic amplitudes, in particular those following
a wide or heavy-tailed distributions (e.g. \citep{Song05_0507,Ikegaya13_293},
reviewed in \citep{buzsaki14_264}), we expect the simple mean-field
approximation applied here to require corrections due to the strong
effect of single synapses.

The generic feature of neuronal dynamics, the threshold-like nonlinearity
that determines the activation of a neuron, is shared by the binary,
the leaky integrate-and-fire and, approximately, also the Hodgkin-Huxley
model neuron. An important approximation entering our theory is the
linearity of the dynamic response with respect to the perturbation.
We estimate the validity of our theory by comparison to direct simulations.
To estimate the breakdown of this approximation we compare the linear
response to the first non-linear correction. We observe that the second
order harmonics in the considered range of parameters remains as small
as about $10$ percent of the first harmonics. The quadratic contribution
to the transfer properties of the neurons stems from the curvature
of the effective gain function $\varphi$ (\prettyref{eq:Def_phi_pop}).
The linear portion of this gain function, in turn, is controlled by
the amplitude $\sigma$ of the synaptic noise. One therefore expects
a breakdown of the linear approximation as soon as the temporal modulation
of the mean input is of the order of this amplitude. \prettyref{fig:Dependence_on_h}
in S1 text shows that with the parameters $\hext=1$ and $\sigma_{\text{exc},\text{inh}}\approx10$,
used in the plots \prettyref{fig:Biological_meanact_frequency}, \prettyref{fig:Biological_c_ee_three_harm}
and  \prettyref{fig:C_ii_three_harm} and \prettyref{fig:C_ei_three_harm},
the linear approximation is good, whereas in \prettyref{fig:Rate_dynamics-1},
we used $\hext=6$, for which the linear perturbation theory already
begins to break down. The latter figure is mainly supposed to give
an intuitive impression. 

A generic property that is shared by nearly all neuron models is the
characteristic duration $\tau$ during which the activity of a sending
cell affects the downstream neuron. For the binary neuron model, this
time scale is identical to the mean interval $\tau$ between updates,
because, once active, a neuron will stay active until the next update.
It most certainly deactivates at that point, because we here consider
low activity states prevalent in cortex \citep{Softky93}. In the
leaky integrate-and-fire model the exponentially decaying membrane
voltage with time constant $\tau$ is qualitatively similar: it sustains
the effect that an input has on the output for this time scale. As
a consequence, neurons transmit their input in a low-pass filtered
manner to their output. This feature persists for more realistic spiking
models, as shown for the leaky integrate-and-fire model \citep{Brunel99,Lindner01_2934},
the exponential integrate-and-fire model \citep{Brunel99,Lindner01_2934},
and the quadratic integrate-and-fire model \citep{Brunel03c}. We
therefore expect that the qualitative properties reported here will
carry over to these models.

A possible application of the framework developed in this paper is
a quantitative comparison of the neuronal activity in the model network
to the analysis of data measured in cortex \citep{Denker11_2681}.
Detecting the occurrence of so called Unitary Events (UE, \citep{Gruen93b,Gruen02a,Gruen02b},
see also  \nameref{subsec:Application-of-the}), the authors observed
that the simultaneous activation of neurons above the level expected
for independence is locked to certain phases of the LFP. They hypothesized
that the reason for this observation is the activation of cell assemblies.
The results presented here show that the correlated activation of
pairs of neurons is modulated by a sinusoidal drive even in a completely
unstructured random network. In consequence, the locking of pairwise
events to the cycle of the LFP is more pronounced for correlated events
than for single spikes. Future work needs to quantitatively compare
experimental data to the results from the model presented here. The
closed form expressions for the modulations of the mean activities
and covariances enable such an approach and the effective study of
the dependence on the model parameters. A quantitative comparison
needs to convert mean activities and pairwise covariances for binary
neurons into the probability to measure a unitary event, interpreting
the binary neuron states as binned spike trains. Preliminary results
indicate that already the homogeneous network presented in this work
can show some features described in \citep{Denker11_2681}. In the
section  \nameref{subsec:Application-of-the}, we apply the Unitary
Event analysis to our setting. The presented methods will be helpful
to analyze the modulation of synchrony in the presence of cell assemblies
\citep{Yoshimura05_868} in the model. This can be done by enhancing
the connection probability among groups of excitatory neurons, similar
as in \citep{LitvinKumar12_e1002667} and will yield a more realistic
model, which captures also nonlinear effects in the perturbation.
Technically this extension amounts to the introduction of additional
populations and the change of the connectivity matrix to reflect that
these populations represent cell assemblies.

The relation of spiking activity to mesoscopic measures, such as the
LFP, is still an open question. These population measures of neuronal
activity naturally depend on the statistics of the microscopic activity
they are composed of. Pairwise covariances, the focus of the current
work, in particular tend to dominate the variance of any mesoscopic
signal of summed activity: The contribution of covariances grows quadratically
in the number of components, the contribution of variances only linearly
\citep[Box 2]{Harris11_509}\citep[eq. (1)]{Tetzlaff12_e1002596}\citep[eq. (1),(2)]{Linden11_859}.
Under the assumption that the LFP mainly reflects the input to a local
recurrent network \citep{Linden11_859,Mazzoni2015}, we have shown
here that these two signals - spikes and LFPs - are intimately related;
not only does the afferent oscillatory drive trivially modulate the
propensity to produce spikes, their firing rate, but also the joint
statistics of pairs of neurons by the three distinct pathways exposed
in the present analysis. Forward modeling studies have shown that
the spatial reach of the LFP critically depends on covariances, with
elevated covariances leading to larger reach \citep{Linden11_859}.
In this light our work shows that a local piece of neuronal tissue
driven by a source of coherent oscillations will more effectively
contribute to the local field potential itself: not only the spiking
rate is modulated accordingly, but also the covariances are increased
and decreased in a periodic manner, further amplifying the modulation
of the generated local field potential and temporally modulating the
spatial reach of the signal.

Functional consequences of the findings presented here deduce from
the hypothesis that communication channels in cortex may effectively
be multiplexed by the selective excitation of different areas with
coherent oscillations \citep{Singer99_49,Womelsdorf07_1609}. The
presented analysis exposes that oscillatory drive to a local piece
of cortex alone already effectively enhances coherent firing beyond
the level expected based on the assumption of independence. If synchronous
activity is employed as a dimension to represent information, it is
hence tightly entangled with time-dependent changes of the mean activity.
A similar conclusion was drawn from the observation that covariance
transmission in feed-forward networks is monotonously increasing with
firing rate \citep{DeLaRocha07_802,Shea-Brown08}. Any information-carrying
modulation of synchronous activity must hence go beyond the here investigated
effects, which can be regarded the baseline given by the non-stationary
activity in networks without function. Since the mechanisms we have
exposed only depend on generic features of cortical tissue - networks
of non-linear neurons, connectivity with strong convergence and divergence,
and dynamic stabilization by inhibition - the time-dependent entanglement
of mean activity and covariances qualitatively exists in any network
with these properties. In this view, our analysis can help to distinguish
the level of time-modulated covariances in neural tissues that are
surprising, and are therefore candidates to be attributed to function,
from those that need to be expected in networks due to their generic
properties.

\section*{Acknowledgement}

The authors would like to thank PierGianLuca Porta Mana for his great
support and Michael Denker, Sonja Gr\"{u}n and the whole INM-6 for
fruitful discussions. All network simulations were carried out with
NEST (http://www.nest-simulator.org).

\newpage{}

\newpage

\section*{Appendix}

\subsection*{Application of the Unitary Event Analysis to correlated network activity\label{subsec:Application-of-the}}
\begin{center}
\textcolor{black}{}
\begin{figure}[H]
\textcolor{black}{\includegraphics[scale=0.75]{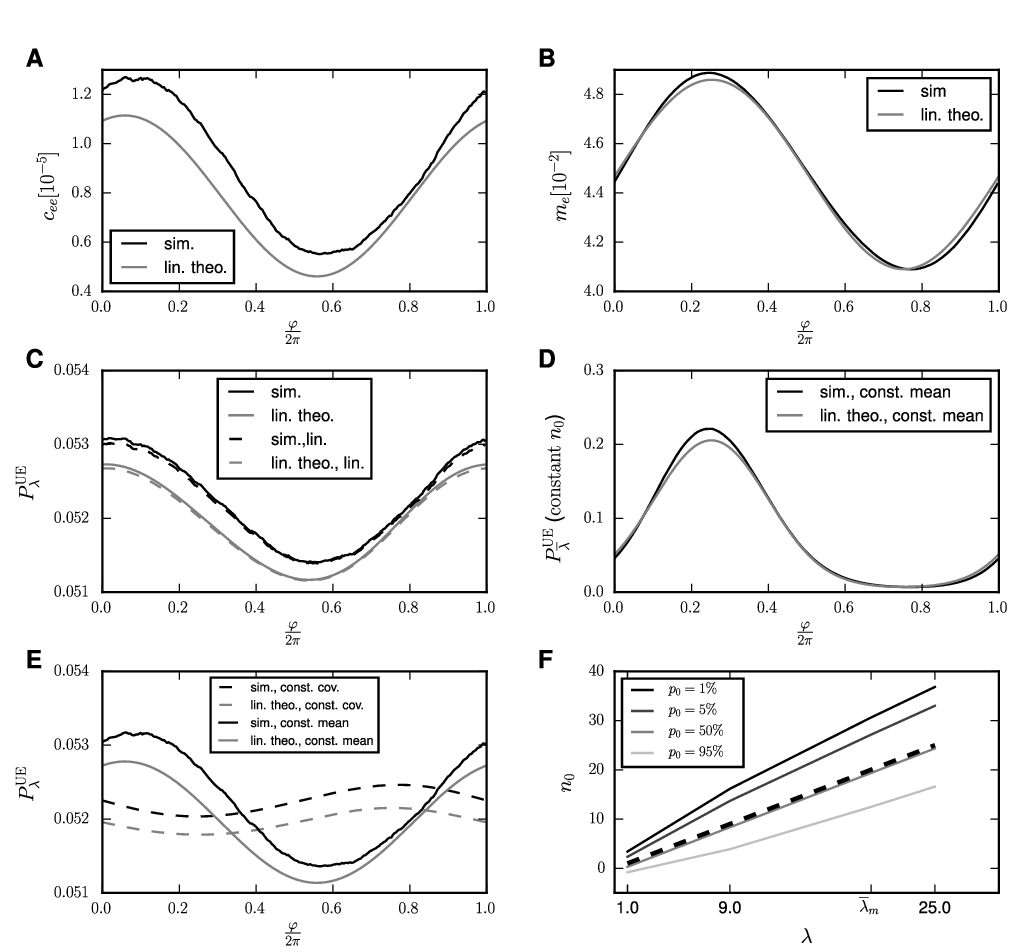}\caption{\textcolor{black}{Temporal modulation of Unitary Events. Covariance
$c(\varphi)$ (A) mean activity $m(\varphi)$ (B) as functions of
the phase $\varphi$ of the LFP cycle. C Probability $P_{\lambda}^{\mathrm{UE}}(\varphi)$
for the appearance of a significant number of Unitary Events as a
function of the phase of the oscillation. Solid curves show the exact
expression \prettyref{appeq:UE_prob_exact}, dashed curves the corresponding
approximation to linear order in $c$, \prettyref{appeq:UE_prob_Taylor}.
D $P_{\lambda}^{\mathrm{UE}}(\varphi)$ for a constant $n_{0}$, adjusted
to the time-averaged mean activity. E $P_{\lambda}^{\mathrm{UE}}(\varphi)$
for constant mean activity and time-dependent covariance (solid curves)
and vice versa (dashed curves). For the plots in A-E, black curves
always represent simulation results and gray curves $N_{\mathrm{bin}}=10000$
and $p_{0}=5
$ and $f=160\protect\Hz$. The parameters for the network simulation
are given in \prettyref{tab:Biological_parameters}. F Dependence
of the lowest number required for a UE $n_{0}$ on the average rate
$\lambda$ for different significance level $p_{0}$, $\overline{\lambda}_{m}$
indicates the stationary part of the rate used for the plots A-E.}\label{fig:Temporal_modulation_of_UE}}
}
\end{figure}
\par\end{center}

We will give a concise, but self-contained description of the main
idea of a Unitary Event analysis and its application to our setup.
The observation of at least $n_{0}$ simultaneous spikes in a time
series of $N_{\mathrm{bin}}$ bins is called Unitary Event (UE, \citep{Gruen93b,Gruen02a,Gruen02b}).
We are therefore interested in the the time-dependence of the probability
$p_{ij}(t)$ that the pair of neurons $i$ and $j$ fires together
at time $t$, which causes a time-dependence of the covariance $c(t)$.
Concretely, because the appearance of a spike is a binary event, the
probability of the joint firing is identical to the second moment
$p_{ij}(t)=\langle n_{i}(t)n_{j}(t)\rangle$ \citep[see also ][ eq. 22]{Glauber63_294}
which, in turn, can be expressed as 
\begin{align}
p_{ij}(t) & =\langle n_{i}(t)n_{j}(t)\rangle=c_{ij}(t)+m_{i}(t)\,m_{j}(t).\label{appeq:p_pair_c_m}
\end{align}
The covariance therefore enters this probability in an additive manner.
The significance test of the Unitary Event analysis, depending on
the momentary rate, aims to eliminate the contribution of the trivial
second term. One therefore expects that the modulation of the covariance
influences also the probability to observe a Unitary Event.

Concretely, one assumes that the number of joint firing events is
Poisson distributed, therefore the probability to observe a UE is
given by
\[
P_{\lambda}^{\mathrm{UE}}\left(n_{0}\right)=\sum_{n>n_{0}}e^{-\lambda_{m}}\frac{\lambda_{m}^{n}}{n!},
\]
where $\lambda=m^{2}N_{\mathrm{bin}}$ for a uncorrelated system with
mean activity $m$ and the number of bins $N_{\mathrm{bin}}$ and
$n_{0}$ is chosen minimal such that $P_{\mathrm{UE}}\left(n_{0}\right)<p_{0}$
for a given significance level $p_{0}$, in \citep{Denker11_2681}
for example, $p_{0}=0.05$. In our setup, $m$ changes continuously
in time, thus the limitation $n_{0}\in\mathbb{N}$ is unfavorable.
Therefore, we replace the cumulative Poisson distribution by a cumulative
distribution yielding the same values on $\mathbb{N}$, but being
defined on $\mathbb{R}$. That is fulfilled by
\[
f\left(\lambda,n_{0}\right):=P_{\lambda}^{\mathrm{UE}}\left(n_{0}\right)=\frac{\gamma\left(n_{0}+1,\lambda\right)}{\Gamma\left(n_{0}+1\right)}=:\frac{\int_{0}^{\lambda}t^{n_{0}}e^{-t}dt}{\int_{0}^{\infty}t^{n_{0}}e^{-t}dt}.
\]
$\Gamma\left(n_{0}\right)$ and $\gamma\left(n_{0}\right)$ are the
Gamma- and the incomplete Gamma-function, respectively. This correspondence
follows from the third last equality in \prettyref{appeq:equivalence_incomplete_gamma_poisson}
and $\Gamma\left(n_{0}+1\right)=n_{0}!\ \forall n_{0}\in\mathbb{N}$.
 Here $f$ is monotonous in $n_{0}$, therefore we can define a function
$f^{-1}\left(\lambda,p_{0}\right)$ via
\[
f\left(\lambda,f^{-1}\left(\lambda,p_{0}\right)\right)=p_{0}.
\]
Now, we want to determine the probability to observe a UE in a correlated
system, that is $\lambda=\lambda_{m}+\lambda_{c}=:\left(m^{2}+c\right)N_{\mathrm{bin}}$
in case that $n_{0}$ is determined assuming a uncorrelated system.
For the systems described in this work, being in the balanced state,
we can safely assume that the covariance $c$ is small and therefore
enters in $P_{\lambda}^{\mathrm{UE}}$ only in linear order:
\begin{align}
 & f\left(\lambda_{m}+\lambda_{c},f^{-1}\left(\lambda_{m},p_{0}\right)\right)\label{appeq:UE_prob_exact}\\
= & f\left(\lambda_{m},f^{-1}\left(\lambda_{m},p_{0}\right)\right)+\partial_{1}f\left(\lambda_{m},f^{-1}\left(\lambda_{m},p_{0}\right)\right)\lambda_{c}+\mathcal{O}\left(\lambda_{c}^{2}\right)\nonumber \\
= & p_{0}+\partial_{1}f\left(\lambda_{m},f^{-1}\left(\lambda_{m},p_{0}\right)\right)\lambda_{c}+\mathcal{O}\left(\lambda_{c}^{2}\right)\label{appeq:UE_prob_Taylor}
\end{align}
($\partial_{1}f$ means the derivative of $f$ with respect to its
first argument). The following computation\foreignlanguage{english}{
\begin{align}
 & P_{\lambda}^{\mathrm{UE}}\left(n_{0}>X\geq n_{0}-1\right)\nonumber \\
= & P_{\lambda}^{\mathrm{UE}}\left(X\geq n_{0}-1\right)-P_{\lambda}^{\mathrm{UE}}\left(X\geq n_{0}\right)\nonumber \\
= & \frac{\int_{0}^{\lambda}t^{n_{0}-1}e^{-t}dt}{\int_{0}^{\infty}t^{n_{0}-1}e^{-t}dt}-\frac{\int_{0}^{\lambda}t^{n_{0}}e^{-t}dt}{\int_{0}^{\infty}t^{n_{0}}e^{-t}dt}\nonumber \\
\overset{\mathrm{P.I.}}{=} & \frac{\lambda^{n_{0}}e^{-\lambda}}{\int_{0}^{\infty}t^{n_{0}}e^{-t}dt}=\frac{\partial}{\partial\lambda}P_{\lambda}\left(X\geq n_{0}\right)\label{appeq:equivalence_incomplete_gamma_poisson}\\
= & \partial_{1}f\left(\lambda_{m},f^{-1}\left(\lambda_{m},p_{0}\right)\right)\nonumber 
\end{align}
}leads to an illustrative interpretation of \prettyref{appeq:UE_prob_Taylor}:
For $c>0$, $\lambda_{c}$ is the number of additional joint firing
events that one expects due to the positive covariance and $\partial_{1}f$
is the probability to observe one joint firing event less than the
minimal number $n_{0}$, that is required for a UE in the uncorrelated
system. Therefore, in this approximation, the required number of joint
firing events for the classification as UE stays the same, only the
probability to observe this many joint firing events is elevated by
$P_{\lambda}^{\mathrm{UE}}\left(s>X\geq s-1\right)\lambda_{c}$.

If we neglect any time-dependence and determine just a constant $n_{0}$
according to the time-averaged mean activity, $P_{\lambda}^{\mathrm{UE}}$
is misestimated for our network (\prettyref{fig:Temporal_modulation_of_UE},
D). Determining $n_{0}$, we therefore have to consider the time-dependence
of $m$. To this end, we can assume that the time-varying part is
small compared to the stationary part, that is
\begin{align*}
\lambda & =\overline{\lambda}_{m}+\delta\lambda_{m}\left(t\right)+\overline{\lambda}_{c}+\delta\lambda_{c}\left(t\right)\\
 & =\left(\overline{m}^{2}+2\overline{m}\delta m\left(t\right)+\overline{c}+\delta c\left(t\right)+\mathcal{O}\left(\delta m\right)^{2}\right)T.
\end{align*}

The qualitative effect of a time-dependent mean activity (which causes
a instantaneous shift in $n_{0}$) in a network with constant positive
covariance can now be seen by the following argument: Assume that
one could instantaneously adjust the covariance such that $n_{0}\left(t\right)=\lambda_{m}\left(t\right)+\lambda_{c}\left(t\right)$,
that is, we construct a system that produces \textit{on average }the
number of joint firing events required \textit{at the minimum} to
be classified as a UE. Like that, the surprise of an observer knowing
this covariance is always on the same level. Following this construction,
a small deviation in $\lambda_{m}\left(t\right)$ around some stationary
value $\overline{\lambda}_{m}$ will force us to also shift $\lambda_{c}\left(t\right)$
a bit according to
\[
\delta\lambda_{c}\left(t\right)=\left(\frac{\partial n}{\partial\lambda}-1\right)\delta\lambda_{m}\left(t\right).
\]
From \prettyref{fig:Temporal_modulation_of_UE}F, we can read off
that $\frac{\partial n}{\partial\lambda}>1$ for small $p_{0}$. We
therefore need $\delta\lambda_{c}\left(t\right)$ to modulate in phase
with $\delta\lambda_{m}\left(t\right)$ to keep the surprise constant.
In turn keeping $\lambda_{c}$ constant will lower the probability
for a UE, if $\lambda_{m}$ is raised. This argument explains that
the UE-probability assuming constant, nonzero covariance modulates
in antiphase with $m\left(t\right)$, as shown by the dashed curves
in panel E. The solid lines in the same panel show that for a constant
mean activity, $P_{\lambda}^{\mathrm{UE}}$ modulates proportional
to $\lambda_{c}\left(t\right)$ (or $c\left(t\right)$, respectively),
as expected from \prettyref{appeq:UE_prob_Taylor}. The actual UE-probability,
shown in C, is a superposition of both effects. The comparison to
the linear approximation, shown by the dashed curves, reveals that
neglecting higher order contributions of $\lambda_{c}$ is indeed
appropriate. As expected from \prettyref{appeq:p_pair_c_m}, the probability
of Unitary Events is elevated because the covariance is positive.
As the time-dependent part of the covariance itself is dominated by
the linear response, we overall get a dominating first harmonic in
the modulation of $P_{\mathrm{UE}}(t)$. As a consequence, we cannot
obtain a locking that is strongly localized at a certain phase of
the LFP, in contrast to the experimental observation (cf. Fig 6 of
\citep{Denker11_2681}).

A quantitative examination would require a Taylor expansion of $\partial_{1}f\left(\lambda_{m},f^{-1}\left(\lambda_{m},p_{0}\right)\right)$
in $\delta\lambda_{m}\left(t\right)$, which gives two contributions
with different signs. The first one is positive and arises because
$\delta\lambda_{m}>0$ causes a rise in $P_{\lambda}^{\mathrm{UE}}\left(n_{0}>X\geq n_{0}-1\right)$
for $n_{0}$ kept constant, the second is negative and comes up because
$\delta\lambda_{n}>0$ causes a positive shift in $n_{0}$ which lowers
$P_{\lambda}^{\mathrm{UE}}\left(n_{0}>X\geq n_{0}-1\right)$ for $\lambda_{m}$
kept constant. Numerical checks seem to show that the last contribution
is dominant for the interesting parameter range leading to $\frac{d}{d\lambda_{m}}\partial_{1}f\left(\lambda_{m},f^{-1}\left(\lambda_{m},p_{0}\right)\right)<0$,
as expected because of the qualitative argument given before.

\subsection*{Some theoretical and technical details\label{sec:Appendix_technical_details}}

\subsubsection*{Derivation of the moment equations using the Master equation\label{subsec:Derivation_of_moment_ODEs}}

For completeness, we here derive the differential equations equations
for the first and second moments \prettyref{eq:ODE_first_two_moments},
following previous work \citep{Renart10_587,Helias14,Ginzburg94,Glauber63_294,Buice09_377}.

We multiply the Master equation by $n_{k}$ or $n_{l}n_{k}$ respectively
and get

\begin{align*}
\tau\frac{d}{dt}\left\langle n_{k}\right\rangle \left(t\right)= & \sum_{\boldsymbol{n}\in\left\{ 0,1\right\} ^{N}}\frac{d}{dt}p\left(\boldsymbol{n},t\right)n_{k}=\sum_{\boldsymbol{n}\backslash n_{k}}n_{l}\phi_{k}\left(\boldsymbol{n}\backslash n_{k},t\right)\\
= & \sum_{\boldsymbol{n}\in\left\{ 0,1\right\} ^{N}}n_{k}\sum\left(2n_{i}-1\right)\phi_{i}\left(\boldsymbol{n}\backslash n_{i},t\right)\\
= & \sum_{\boldsymbol{n}\in\left\{ 0,1\right\} ^{N}}\left(n_{k}\phi\left(\boldsymbol{n}\backslash n_{k},t\right)+n_{k}\underbrace{\sum_{i\neq k}^{N}\left(2n_{i}-1\right)\phi_{i}\left(\boldsymbol{n}\backslash n_{i},t\right)}_{=0}\right)\\
= & \sum_{\boldsymbol{n}\backslash n_{k}}\left[-p\left(\boldsymbol{n}_{k+},t\right)+\left(p\left(\boldsymbol{n}_{k-},t\right)F_{k}\left(\boldsymbol{n}_{k-}\right)+p\left(\boldsymbol{n}_{k+},t\right)F_{k}\left(\boldsymbol{n}_{k+}\right)\right)\right]\\
= & -\left\langle n_{k}\right\rangle \left(t\right)+\left\langle F_{k}\left(t\right)\right\rangle 
\end{align*}

and
\begin{eqnarray*}
\frac{d}{dt}\left\langle n_{k}\left(t\right)n_{l}\left(t\right)\right\rangle  & = & \sum_{\boldsymbol{n}\in\left\{ 0,1\right\} ^{N}}\frac{d}{dt}p\left(\boldsymbol{n},t\right)n_{k}n_{l}\\
 & = & \sum_{\boldsymbol{n}\in\left\{ 0,1\right\} ^{N}}n_{k}n_{l}\sum_{i=1}^{N}\left(2n_{i}-1\right)\phi_{i}\left(\boldsymbol{n}\backslash n_{i},t\right)\\
 & = & \sum_{\boldsymbol{n}\in\left\{ 0,1\right\} ^{N}}\left(n_{k}n_{l}\phi_{k}\left(\boldsymbol{n}\backslash n_{k},t\right)+n_{l}n_{k}\phi_{l}\left(\boldsymbol{n}\backslash n_{l},t\right)\right.\\
 & + & n_{k}n_{l}\underbrace{\sum_{i\neq k,l}^{N}\left(2n_{i}-1\right)\phi_{i}\left(\boldsymbol{n}\backslash n_{i},t\right)}_{=0}\left.\right)\\
 & = & \sum_{\boldsymbol{n}\backslash n_{k}}n_{l}\phi_{k}\left(\boldsymbol{n}\backslash n_{k},t\right)+k\leftrightarrow l\\
 & = & \sum_{\boldsymbol{n}\backslash n_{k}}\left[-n_{l}p\left(\boldsymbol{n}_{k+},t\right)+n_{l}\left(p\left(\boldsymbol{n}_{k-},t\right)F_{k}\left(\boldsymbol{n}_{k-}\right)+p\left(\boldsymbol{n}_{k+},t\right)F_{k}\left(\boldsymbol{n}_{k+}\right)\right)\right]\\
 &  & +k\leftrightarrow l\\
 & = & \left\{ -\left\langle n_{k}\left(t\right)n_{l}\left(t\right)\right\rangle +\left\langle n_{l}\left(t\right)F_{k}\left(t\right)\right\rangle \right\} +\left\{ k\leftrightarrow l\right\} .
\end{eqnarray*}

\subsubsection*{}
\selectlanguage{english}%

\subsubsection*{}
\selectlanguage{american}%

\subsubsection*{Different definitions for a spiking event of a binary neuron\label{subsec:Spike_defs}}

In \citep{VanVreeswijk98_1321}, van Vreeswijk et al. identify the
transition $0\rightarrow1$ with a spike, which leads to the equation
$\nu_{\alpha}=\frac{m_{\alpha}\left(1-m_{\alpha}\right)}{\tau}$ for
the firing rate. We think, however, that this identification is inappropriate
in our case, because the $0\rightarrow1$-transition for a binary
neuron has a different meaning than a spike for a spiking neuron.
In our opinion, it is decisive, for which fraction of time a spiking
neuron affects the downstream neurons. If it spikes with frequency
$\nu_{\alpha}$ and the membrane potential decays with the time constant
$\tau$, this fraction is given by $\tau\nu_{\alpha}$. This can be
interpreted as the mean activity of a spiking neuron, which leads
to the definition of the firing rate of a binary neuron $\nu_{\alpha}=\frac{m_{\alpha}}{\tau}$
in section \nameref{subsec:Two-populations-with}. In other words:
If we want to identify a spiking event for a binary neuron, we will
have to count the $1\rightarrow1$-transition as spike as well. For
small mean activities, however, the difference is small anyway. 

\subsubsection*{Extracting the correct phase from complex solutions\label{subsec:Phase_extraction}}

Notice that there are a few subtleties to keep in mind when a discrete
Fourier transform is applied to $\delta m_{\alpha}$. The (in both
senses) real-valued solution of the ODE is
\begin{align*}
\delta m_{\alpha} & =\Im\left(M_{\alpha}^{1}e^{i\omega_{0}t}\right)=\Im\left(\left|M_{\alpha}^{1}\right|e^{i\left(\arg\left(M_{\alpha}^{1}\right)+\omega_{0}t\right)}\right)=\left|M_{\alpha}^{1}\right|\sin\left(\arg\left(M_{\alpha}^{1}\right)+\omega_{0}t\right)\\
 & =\left|M_{\alpha}^{1}\right|\left(\sin\left(\arg\left(M_{\alpha}^{1}\right)\right)\cos\left(\omega_{0}t\right)+\cos\left(\arg\left(M_{\alpha}^{1}\right)\right)\sin\left(\omega_{0}t\right)\right).
\end{align*}
For clarity, we here named the driving frequency $\omega_{0}$. Therefore,
if we calculate the Fourier transform (in a distributional sense),
we get
\begin{align*}
 & {\cal F}\left[\delta m_{\alpha}\right]\left(\omega=\omega_{0}\right)\\
= & \left|M_{\alpha}^{1}\right|\left(\sin\left(\arg\left(M_{\alpha}^{1}\right)\right)\frac{\delta_{\omega_{0}}+\delta_{-\omega_{0}}}{2}+\cos\left(\arg\left(M_{\alpha}^{1}\right)\right)\frac{\delta_{\omega_{0}}-\delta_{-\omega_{0}}}{2i}\right)\\
= & \frac{\left|M_{\alpha}^{1}\right|}{2}\left(\delta_{\omega_{0}}\left(\sin\left(\arg\left(M_{\alpha}^{1}\right)\right)-i\cos\left(\arg\left(M_{\alpha}^{1}\right)\right)\right)+\delta_{-\omega_{0}}\left(\sin\left(\arg\left(M_{\alpha}^{1}\right)\right)+i\cos\left(\arg\left(M_{\alpha}^{1}\right)\right)\right)\right)
\end{align*}
Thus, we get
\[
\left|{\cal F}\left[\delta m_{\alpha}\right]\left(\omega_{0}\right)\right|=\left|{\cal F}\left[\delta m_{\alpha}\right]\left(-\omega_{0}\right)\right|=\frac{\left|M_{\alpha}^{1}\right|}{2}
\]
and, because we take the imaginary part of the complex solution which
leads to a $\frac{\pi}{2}$-phase shift compared to the complex phase\foreignlanguage{english}{
\[
\arg\left({\cal F}\left[\delta m_{\alpha}\right]\left(\omega_{0}\right)\right)=\begin{cases}
\text{arg}\left(M_{\alpha}^{1}\right)+\frac{3\pi}{2}, & \text{for }\text{arg}\left(M_{\alpha}^{1}\right)\in\left[-\pi,-\frac{\pi}{2}\right]\\
\text{arg}\left(M_{\alpha}^{1}\right)-\frac{\pi}{2}, & \text{for }\text{arg}\left(M_{\alpha}^{1}\right)\in\left(-\frac{\pi}{2},\pi\right).
\end{cases}
\]
}

\subsection*{Comparison of simulation and theory of the EI and II-covariances
and validation of the linear perturbation theory}

For completeness, we include here the plots showing the dependence
of the covariances between inhibitory and inhibitory and excitatory
on the driving frequency for the third network setup of the main text.
In \prettyref{fig:Dependence_on_h}, we show that the linear perturbation
theory breaks down if the perturbation is of the same order as the
input fluctuations.

\selectlanguage{english}%
\begin{figure}[h]
\begin{centering}
\includegraphics[scale=0.65]{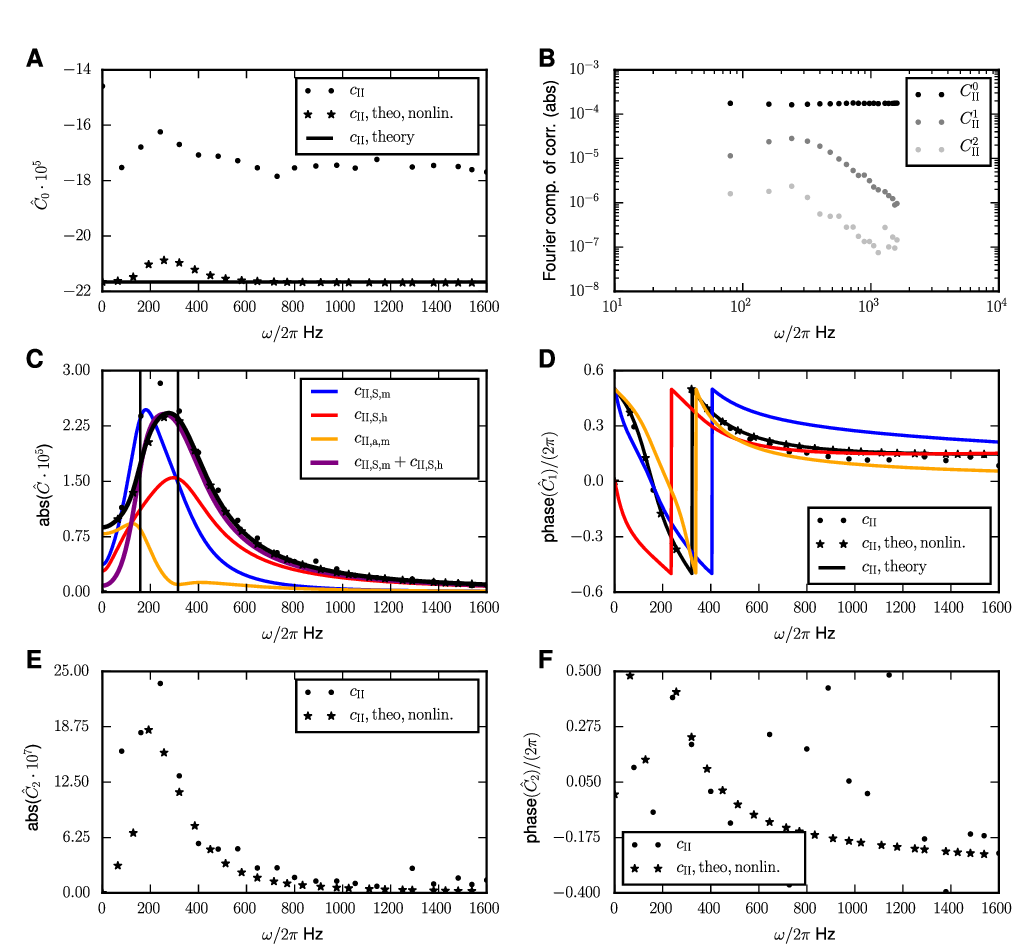}
\par\end{centering}
\caption{\foreignlanguage{american}{\textbf{Driven E-I network with biologically inspired parameters:
II-Covariance.} Response of the inh.-inh.-part of the covariance to
a perturbation with frequency $\omega$ in the Fourier space.\textbf{
A} Zeroth order (time independent part) of the covariance. \textbf{B}
Absolut value of the first three Fourier components of the $c_{\mathrm{II}}$-covariances
in loglog-scale. \textbf{C} Absolute value of the first order of the
time-dependent part of the covariance. \textbf{D} Phase angle in relation
to the driving signal. \textbf{E} and \textbf{F} analogous to \textbf{C}
and \textbf{D} for the second Fourier modes. Solid lines indicate
the linear theory \prettyref{eq:Correlation_Fourier}, stars the results
of the numerical solved full mean-field theory \prettyref{eq:ODE_mean_population_averaged}
and \prettyref{eq:ODE_correlation_population_averaged} and dots those
of the direct simulation of the full network\foreignlanguage{english}{.}
Numerical results obtained by the same methods and identical parameters
as in \prettyref{fig:Biological_meanact_frequency}.\foreignlanguage{english}{\label{fig:C_ii_three_harm}}}}
\end{figure}

\begin{figure}[h]
\centering{}\includegraphics[scale=0.65]{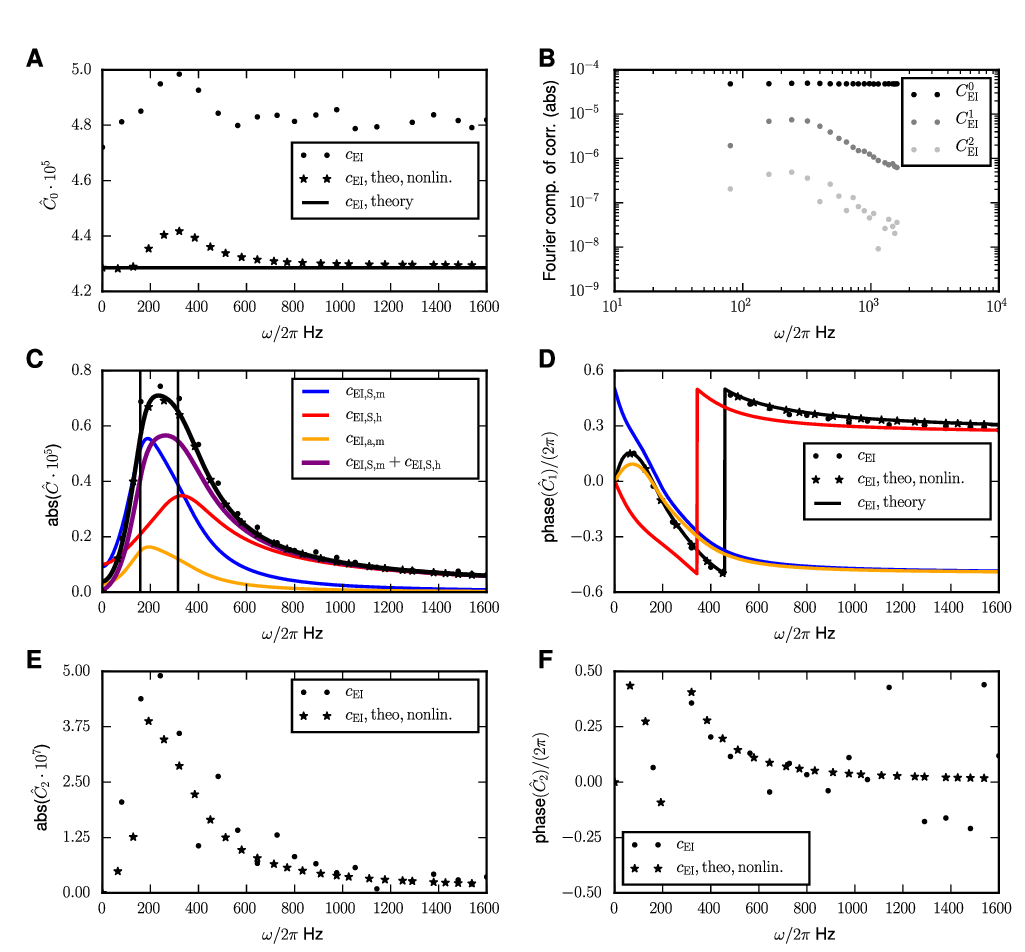}\caption{\foreignlanguage{american}{\textbf{Driven E-I network with biologically inspired parameters:
EI-Covariance.} Response of the exc.-inh.-part of the covariance to
a perturbation with frequency $\omega$ in the Fourier space.\textbf{
A} Zeroth order (time independent part) of the covariance.\textbf{
B} Absolut value of the first three Fourier components of the $c_{\mathrm{EI}}$-covariances
in loglog-scale. \textbf{C} Absolute value of the first order of the
time-dependent part of the covariance. \textbf{D} Phase angle in relation
to the driving signal. \textbf{E} and \textbf{F} analogous to \textbf{C}
and \textbf{D} for the second Fourier modes. Solid lines indicate
the linear theory \prettyref{eq:Correlation_Fourier}, stars the results
of the numerical solved full mean-field theory \prettyref{eq:ODE_mean_population_averaged}
and \prettyref{eq:ODE_correlation_population_averaged} and dots those
of the direct simulation of the full network\foreignlanguage{english}{.}
Numerical results obtained by the same methods and identical parameters
as in \prettyref{fig:Biological_meanact_frequency}.\foreignlanguage{english}{\label{fig:C_ei_three_harm}}}}
\end{figure}

\begin{figure}[h]
\selectlanguage{american}%
\begin{centering}
\includegraphics[scale=0.75]{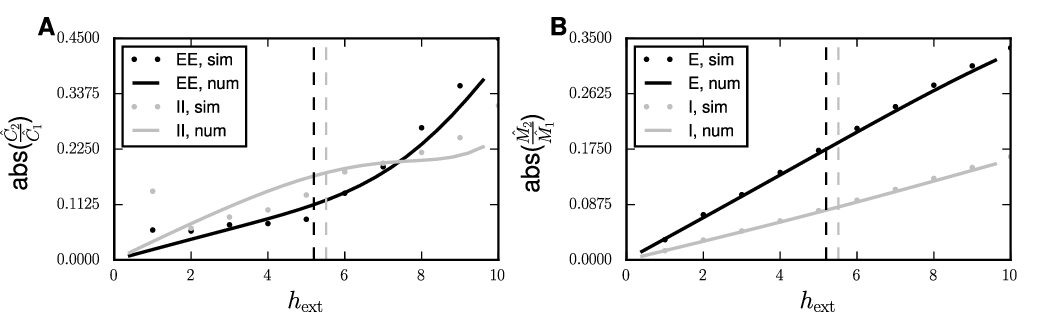}
\par\end{centering}
\selectlanguage{english}%
\caption{\foreignlanguage{american}{\textbf{Driven E-I network with biologically inspired parameters:
Dependence of the covariance and the mean activity on $\protect\hext$}.
Ratio of the second to the first Fourier component in a system subject
to a perturbation with frequency $\omega=20\cdot2\pi\mathrm{Hz}$.\textbf{
A} Covariance between excitatory and between inhibitory neurons. \textbf{B}
Mean activity of the excitatory and of the inhibitory population.
The vertical dotted lines indicate $\sigma_{\mathrm{exc.}}/2$ (black)
and $\sigma_{\mathrm{inh.}}/2$ (lightgray). Solid lines indicate
the results of the numerical solved full mean-field theory \prettyref{eq:ODE_mean_population_averaged}
and \prettyref{eq:ODE_correlation_population_averaged} and dots those
of the direct simulation of the full network\foreignlanguage{english}{.}
Numerical results obtained by the same methods and with the same parameters
as in \prettyref{fig:Single_pop_frequency}.\label{fig:Dependence_on_h} }}
\end{figure}

\selectlanguage{american}%

\section*{}

\subsection*{}

\end{document}